\documentclass[aps,prb,twocolumn,showpacs,floatfix,pdf]{revtex4}

\usepackage{amsmath}    
\usepackage{graphicx}   

\begin{document}
\title{
Advanced multi-orbital impurity solver for dynamical mean field theory based on the equation of motion approach}

\author{Qingguo Feng}
\email[]{qingguo.feng@physics.uu.se}
\affiliation{Department of Physics and Astronomy, Uppsala University, Box 516, S-75120 Uppsala, Sweden}
\author{P. M. Oppeneer}
\affiliation{Department of Physics and Astronomy, Uppsala University, Box 516, S-75120 Uppsala, Sweden}

\date{\today}

\begin{abstract}
We propose an improved fast multi-orbital impurity solver for the dynamical mean field theory (DMFT) based on equations of motion (EOM) of Green's functions and decoupling scheme.
In this scheme the inter-orbital Coulomb interactions are treated fully self-consistently, involving the inter-orbital fluctuations.
As an example of the derived multi-orbital impurity solver, the two-orbital Hubbard model is studied for various cases. Comparisons are made between numerical results obtained with our EOM scheme and those obtained with quantum Monte Carlo and numerical renormalization group  methods. The comparison substantiates a good agreement, but also reveals a dissimilar behavior in the densities of states (DOS) which is caused by inter-site inter-orbital hopping effects and on-site inter-orbital fluctuation effects,
thus corroborating the value of the EOM method for the study of multi-orbital strongly correlated systems.
\end{abstract}


\maketitle

\section{Introduction\label{sect1}}
The dynamical mean field theory (DMFT), which was proposed \cite{DMFT1,DMFT2} and developed in the past decade,
 is presently one of the most widely used methods for studying strongly correlated electron systems.  Although it is exact only for infinite dimensional systems, it has been proven that it can provide a good approximation even for three dimensional systems. For low dimensional systems, a cellular DMFT has been developed in order to take into account spacial fluctuations. For a review of DMFT, see Refs.\ \onlinecite{DMFTRMP96} and \onlinecite{DMFTRMP06}.

In the DMFT, the main tasks are to solve the so-called impurity problem and to derive the DMFT self-consistency conditions for a specified lattice of a real material. For the former task, several  numerical techniques have been developed to construct the impurity solver, e.g., the quantum Monte Carlo (QMC) method,\cite{HF-QMC} the exact diagonalization (ED) method,\cite{ED,ED2} numerical renormalization group (NRG) method,\cite{NRG,NRG2} density matrix renormalization group (DMRG) method,\cite{DMRG} the continuous time quantum Monte Carlo (CTQMC) method,\cite{CTQMC,CTQMC2} and the equation of motion (EOM) method.\cite{JK05,FZJ09,FO11}
The EOM method, which is the focus of the present article, works directly on the real frequency axis and for any temperature, and  in addition has numerical advantages as it is computationally faster than other numerical methods, because other methods are all computationally expensive due to the involved matrix manipulations, in which the Hilbert space dimension increases exponentially with the number of orbitals. Moreover, the EOM method is a very general method that can be implemented to a large variety of systems in different cases. Also, through the study of equations of motions, we can learn more about the physics underlying interesting many-body systems, and do not only focus on the numerical technique and code development. The EOM method has actually been used by J. Hubbard in his corner stone work\cite{Hubbard63,Hubbard64a,Hubbard64b} and has continually been studied by various people in studies of the single-impurity Anderson model (SIAM) and Hubbard model
(see,  e.g., Refs.\ \onlinecite{Lacroix1, Lacroix2,Czycholl85,Petru93,GrosEOM,Kang95,Luo99,ZhuEOM1}). Here we study the impurity solver based on the EOM method within the DMFT framework.
This study will also be beneficial for the development of further techniques, for example cellular DMFT for low-dimension systems beyond the standard DMFT,  or development of techniques for directly solving the multi-orbital lattice model with tight binding.

On the basis of equations of motion and decoupling techniques,
an infinite $U$ single orbital impurity solver has been developed in Ref.~\onlinecite{JK05} and a finite $U$ single orbital impurity solver with spin-orbital degeneracy $N$ has been developed in Ref.~\onlinecite{FZJ09}.
Recently, a multi-orbital equation-of-motion (MO-EOM) impurity solver has been developed by us
 in Ref.~\onlinecite{FO11}, which correctly captures the physics of many-body systems with the inclusion of the inter-site multi-orbital hopping interactions.
 However, in Ref.~\onlinecite{FO11} we have used the mean-field approximation for the on-site inter-orbital Coulomb interactions and have neglected the inter-orbital fluctuations, which is approximately valid for those systems where the inter-orbital effects are weak or there exists strong screening. For those systems with strong inter-orbital fluctuations, which can happen, e.g., when the orbital levels are energetically very close or when an applied external field exists, the inter-orbital fluctuations can be important and have to be included.

In this paper, we have developed an improved EOM method for strongly correlated multi-orbital many electron systems, in which we have implemented a higher-order decoupling scheme, where the on-site inter-orbital fluctuations associated with the inter-orbital Coulomb interactions are well included. A double check of the numerical results and iterative convergence procedure are introduced by calculating both the electron system and the associated `hole' quasi-particle system in order to obtain accurate results. Using this MO-EOM scheme we find  that the Kondo peak at the Fermi level is well reproduced for heavy fermions, and also the Mott metal-insulator transition \cite{ImadaRMP1998} is observed. In addition, it is observed that the Hubbard bands are splitted due to the different intra-orbital and inter-orbital Coulomb interaction strengths, and the quasiparticle DOS shows therefore an interesting multi-peak structure. This feature has been studied by tracing the difference between a two-orbital system with identical band widths for the two orbitals and that with different band widths.  Comparisons are performed between results obtained with our MO-EOM method and those obtained with NRG and QMC methods.

The paper is organized as follows: The equations of motion, decoupling scheme, and physical assumptions are introduced in Sect.~\ref{sect2}. Then in Sect.~\ref{sect3}, the results for the two-orbital Hubbard model as an example are discussed. Finally, a summary of the work is given in Sect.~\ref{sect4}.

\section{Description of MO-EOM method\label{sect2}}
We start our description from the model Hamiltonian. In many-body physics, the Hubbard model, in which  only the hopping of electrons between the sites and the on-site Coulomb interactions are considered, is the simplest yet one of the most important lattice models. For a multi-orbital system, the Hamiltonian can be written as
\begin{eqnarray}
{\cal H}&=&-\sum_{ijlm\sigma,i\neq j}t_{ijlm}f^{\dag}_{il\sigma}f_{jm\sigma}
+\sum_{il}U_{ll}\hat{n}_{il\uparrow}\hat{n}_{il\downarrow}\nonumber\\
&&+\sum_{ilm\sigma\sigma',l<m}U_{lm\sigma\sigma'}\hat{n}_{il\sigma}\hat{n}_{im\sigma'}\nonumber\\
&&+\sum_{ilm\sigma,l<m}\big(V'^{\ast}_{lm\sigma}f^{\dag}_{im\sigma}f_{il\sigma}^{~}+V'_{lm\sigma}f^{\dag}_{il\sigma}f^{~}_{im\sigma}\big),\label{eq:1}
\end{eqnarray}
where $i,j$ are the site indices, $l,m$ are the orbital indices, $\sigma, \sigma'$ are the spin indices, and $f^{\dag}_{il\sigma}, f_{il\sigma}$ are the creation and annihilation operators for electrons with $\sigma$ spin in $l$-th orbital, respectively.
The first term describes the hopping of electrons between the different sites. The second (third) term is the on-site intra-orbital (inter-orbital) Coulomb interaction term. The last two terms are the on-site inter-orbital single hopping of electrons, where $V'^{\ast}_{lm\sigma}$ and $V'_{lm\sigma}$ are the inter-orbital hopping amplitudes for spin $\sigma$ between the $l$-th and $m$-th orbitals.

In the dynamical mean field theory, the lattice model is mapped to an impurity model, usually the single-impurity Anderson model (SIAM), and the interactions between sites are mapped to the interactions between the impurity and a bath. For Eq.\ \eqref{eq:1}, the mapped SIAM has the following Hamiltonian,
\begin{eqnarray}
{\cal H}_{imp}&=&\sum_{kl\sigma}\varepsilon_{kl\sigma}c^{\dag}_{l k\sigma}c_{l k\sigma}^{~}+\sum_{l\sigma}\varepsilon_{fl\sigma} f^{\dag}_{l\sigma}f_{l\sigma}+\sum_{l}U_{ll}\hat{n}_{l\uparrow}\hat{n}_{l\downarrow}\nonumber\\
&+&\sum_{lm\sigma\sigma',l<m}U_{lm\sigma\sigma'}\hat{n}_{l\sigma}\hat{n}_{m\sigma'}\nonumber\\
&+&~~\sum_{l k\sigma}~~~~\big(V^{\ast}_{l k\sigma}c^{\dag}_{l k\sigma}f_{l\sigma}+V_{l k\sigma}f^{\dag}_{l\sigma}c_{l k\sigma}\big)\nonumber\\
&+&\sum_{lm\sigma,l<m}~\big(V'^{\ast}_{lm\sigma}f^{\dag}_{m\sigma}f_{l\sigma}+V'_{lm\sigma}f^{\dag}_{l\sigma}f_{m\sigma}\big).
\label{eq:2}
\end{eqnarray}
The first (second) term is the energy of conduction electrons (localized electrons), where the electrons in different
orbitals are labeled with the orbital index $l$. $\hat{n}_{l\sigma}=f^{\dag}_{l\sigma}f_{l\sigma}^{~}$ is the occupation number for localized electrons with spin $\sigma$ in the $l$-th orbital, and $\varepsilon_{fl\sigma}$ is the orbital level. The third term is the on-site intra-orbital Coulomb interaction term, and the fourth summation term is the on-site inter-orbital Coulomb interactions between electrons of the $l$-th orbital and $m$-th orbital. The fifth summation including two terms is the hybridization between the localized electrons and the baths. The last two terms are the on-site inter-orbital single hopping, where the site index has been dropped comparing to Eq.~\eqref{eq:1}.

In our studies, we have used the double time temperature-dependent retarded Green's function in Zubarev notation,~\cite{Zubarev}
\begin{eqnarray}
G_{AB}(t,t')
&=&\ll A(t);B(t')\gg\nonumber\\
&=&-i\Theta(t-t')\langle
[A(t),B(t')]_+\rangle,\label{eq:3}
\end{eqnarray}
where $A(t)$ and $B(t')$ are the Heisenberg operators.
and $\Theta(t-t')$ is the Heavyside function.
Applying the Fourier transform, we have obtained the Green's function in $\omega$ space,
which should satisfy the equations of motion
\begin{eqnarray}
\omega\ll A;B\gg=\langle[A,B]_+\rangle+\ll[A,{\cal H}_{imp}];B\gg.\label{eq:4}
\end{eqnarray}

For a multi-orbital electron system, if we define the following anti-commutation relation for the operators,
\begin{eqnarray}
&&[f_{l\sigma}, f_{m\sigma'}]_+=0, ~[f^{\dag}_{l\sigma},
f^{\dag}_{m\sigma'}]_+=0, \nonumber\\
&&[f^{\dag}_{l\sigma}, f_{m\sigma'}]_+=\delta_{lm}\delta_{\sigma\sigma'},
~[f_{l\sigma}, f^{\dag}_{m\sigma'}]_+=\delta_{lm}\delta_{\sigma\sigma'},\label{eq:5}\nonumber
\end{eqnarray}
we can calculate with Eq.\ \eqref{eq:4} and obtain the first two equations of motion as follows (here we only show it for one orbital, $m$),
\begin{widetext}
\begin{eqnarray}
(\omega+\mu-\varepsilon_{fm\sigma})\ll
f_{m\sigma};f^{\dag}_{m\sigma}\gg
&=&1+U_{mm}\ll\hat{n}_{m\sigma'}f_{m\sigma};f^{\dag}_{m\sigma}\gg\nonumber\\
&+&\sum_{l,l\neq m}\big(U_{lm\sigma\sigma}\ll\hat{n}_{l\sigma}f_{m\sigma};f^{\dag}_{m\sigma}\gg
+U_{lm\sigma'\sigma}\ll\hat{n}_{l\sigma'}f_{m\sigma};f^{\dag}_{m\sigma}\gg\big)\nonumber\\
&+&\sum_{k}~\big(V_{mk\sigma}\ll c_{mk\sigma};f^{\dag}_{m\sigma}\gg-\sum_{l,l\neq m}V'_{lm\sigma}\ll f_{l\sigma};f^{\dag}_{m\sigma}\gg\big) ,
\label{eq:6}\\
(\omega+\mu-\varepsilon_{km\sigma})\ll
c_{mk\sigma};f^{\dag}_{m\sigma}\gg&=&
{\cal V}^{\ast}_{mmk\sigma}\ll
f_{m\sigma};f^{\dag}_{m\sigma}\gg+\sum_{l}{\cal V}^{\ast}_{mlk\sigma}\ll f_{l\sigma};f^{\dag}_{m\sigma}\gg_{l\neq m},
\label{eq:7}
\end{eqnarray}
\end{widetext}
where $\mu$ is the chemical potential. Moreover, we have employed the notation $\sigma'\neq\sigma$ which will also be employed in the following context.

The first term of  Eq.~\eqref{eq:6} (i.e., 1) on the right hand side (RHS) reflects the existence of a particle with spin $\sigma$ in the $m$-th orbital itself.
The second term on the RHS of Eq.~\eqref{eq:6} reflects the fluctuation of the spin $\sigma$ in the $m$-th orbital
accompanied by the spin $\sigma'$ in the $m$-th orbital, which can be considered in another way as a fluctuation of the spin $\sigma$ when spin $\sigma'$ exists.
The third (fourth) term gives the fluctuation of the spin $\sigma$ in the $m$-th orbital accompanied by the spin $\sigma$ ($\sigma'$) in the $l$-th orbital, respectively. 
Furthermore, Eq.~\eqref{eq:7} describes the situation that the electrons hop from the bath to the localized orbital, in which event the electrons can hop to {\it any } of the orbitals of this impurity, while in Eq.~\eqref{eq:6} there is the procedure that electrons hop from the localized orbitals to the bath.
Therefore, we can imagine as a physical picture
that one electron hops from the $m$-th orbital of the impurity to the bath and then hops back to the $m'$-th orbital,
where $m'$ can be identical to or different from $m$.
Consequently, the hybridization parameters ${\cal V}^{\ast}_{mmk\sigma}$ and ${\cal V}^{\ast}_{lmk\sigma}$ label the processes that one electron comes from the $m$-th orbital and return to the $m$-th or $l$-th orbital, respectively. Due to the preservation of the charge, there exists the relation that $V^{\ast}_{mk\sigma}={\cal V}^{\ast}_{mmk\sigma}+\sum_{l,l\neq m}{\cal V}^{\ast}_{lmk\sigma}$.
Note that, because the SIAM is a mapped SIAM and therefore the bath in this mapped SIAM is actually a mapped virtue bath, it indeed corresponds to a real process in the lattice model in which one electron hops from one orbital on the studied site to any other orbital on any other site, which is the physical picture.
One should note that ${\cal V}^{\ast}_{mmk\sigma}$ and ${\cal V}^{\ast}_{lmk\sigma}$ will not appear in the impurity Hamiltonian Eq.~\eqref{eq:2}, see Fig.~1 in Ref.~\onlinecite{FO11}.

From the derivation of Eqs.\ \eqref{eq:6} and \eqref{eq:7}, we can note that, when calculating the equation of motion of one Green's function $\ll A;B\gg$, new Green's functions will appear, which are called higher order Green's functions than $\ll A;B\gg$. If we calculate the equations of motion of these higher order Green's functions, even higher order Green's functions will appear in these newly derived equations of motion and these equations of motion are also called higher order equations of motion having a higher order than those in the previous step. Each Green's function only has one order. Only those Green's functions that first appear and never appeared in previous lower order equations of motion will receive its order by observing in which order equations of motion they first appear. Here the order of the Green's function approximately labels the weight of the interaction associated with this Green's function. Repeating this procedure again and again, more and more equations of motion will be calculated and higher and even higher order Green's functions will appear, which is an infinite procedure. Then a decoupling scheme will be employed to truncate this procedure and approximate those higher order Green's functions (higher than this truncation) with the product of the lower order Green's functions and relative correlation functions. Thus the equations are closed and can be solved. The decoupling scheme gives under which order the interactions are treated exactly and all interactions higher than this order are treated approximately. The higher this order is, the more accurate the scheme is. One can always get satisfactory precision of the results by choosing appropriate decoupling scheme, e.g., to choose one order higher decoupling scheme if the employed one is not sufficient.

In Ref.\ \onlinecite{FO11}, we have treated the on-site inter-orbital Coulomb interactions with the mean-field approximation which may cause a loss of interesting information associated with the inter-orbital fluctuations. In the present treatment we take  the inter-orbital fluctuations fully into account. We have previously noticed that, considering the pre-occupied charge besides the two spins in the two-particle Green's function that we are studying, the Coulomb interaction strength should be modified accordingly to be an effective one.\cite{FO11} The DMFT will correct the shape of bands and the charge occupations, where the latter will cause a change of the Coulomb interactions between correlated electrons. Moreover, having the right Coulomb interactions will shift the bands to right positions, while the band positions and the shape of the bands will change the occupations. This implies that occupations, Coulomb interactions, band positions, and shape of bands have to be determined self-consistently. Now, since the inter-orbital fluctuation terms have here been taken into account, there should be three equations to describe the intra-orbital and inter-orbital Coulomb interaction strength, respectively,
\begin{eqnarray}
U^{eff}_{mm}&=&U_{mm}+\sum_{l, l\neq m} (U_{lm\sigma\sigma}\bar{n}_{l\sigma}+U_{lm\sigma\sigma'}\bar{n}_{l\sigma'}),
\end{eqnarray}
\begin{eqnarray}
U^{eff}_{lm\sigma\sigma}&=&U_{lm\sigma\sigma}+U_{mm}\bar{n}_{m\sigma'}+U_{lm\sigma\sigma'}\bar{n}_{l\sigma'}\nonumber\\
&+&\sum_{l', l' \neq m}(U_{l'm\sigma\sigma}\bar{n}_{l'\sigma}+U_{l'm\sigma\sigma'}\bar{n}_{l'\sigma'}),
\end{eqnarray}
\begin{eqnarray}
U^{eff}_{lm\sigma\sigma'}&=&U_{lm\sigma\sigma'}+U_{mm}\bar{n}_{m\sigma'}+U_{lm\sigma\sigma}\bar{n}_{l\sigma}\nonumber\\
&+&\sum_{l', l' \neq m}(U_{l'm\sigma\sigma}\bar{n}_{l'\sigma}+U_{l'm\sigma\sigma'}\bar{n}_{l'\sigma'}),
\end{eqnarray}
where $l\neq m$, $l'\neq l$ and $l'\neq m$.

As a next step, we derive the equations in a form with the total hybridization functions $\Delta_{m\sigma}$, where for the simplicity we neglect the on-site inter-orbital direct single hoppings as we have done in Ref.\ \onlinecite{FO11} and these on-site inter-orbital direct single hoppings will be treated in a forthcoming work. Therefore, we can obtain the following equation of motion,
\begin{eqnarray}
&&(\omega+\mu-\varepsilon_{fm\sigma}-\Delta_{m\sigma})\times\nonumber\\
&&\ll f_{m\sigma}^{~} ;f^{\dag}_{m\sigma}\gg=1+U^{mm}_{\it eff}\ll\hat{n}_{m\sigma'}f_{m\sigma}^{~};f^{\dag}_{m\sigma}\gg\nonumber\\
&&\qquad\qquad\qquad~~~+\sum_{l,l\neq m 
}\big(U_{lm\sigma\sigma}\ll\hat{n}_{l\sigma}f_{m\sigma};f^{\dag}_{m\sigma}\gg\nonumber\\
&&\qquad\qquad\qquad~~~+U_{lm\sigma\sigma}\ll\hat{n}_{l\sigma'}f_{m\sigma};f^{\dag}_{m\sigma}\gg\big),
\label{eq:8}
\end{eqnarray}
where
\begin{eqnarray}
\Delta_{m\sigma}&=&\Delta_{mm\sigma}+\sum_{l,l\neq m} \Delta_{lm\sigma},\\
\Delta_{mm\sigma}&=&\sum_k\frac{V_{mk\sigma}{\cal V}^{\ast}_{mmk\sigma}}{\omega+\mu-\varepsilon_{mk\sigma}},\\
\Delta_{lm\sigma}&=&\sum_k\frac{V_{mk\sigma}{\cal V}^{\ast}_{lmk\sigma}}{\omega+\mu-\varepsilon_{mk\sigma}}\frac{\ll
f_{l\sigma};f^{\dag}_{m\sigma}\gg}{\ll
f_{m\sigma};f^{\dag}_{m\sigma}\gg}.
\label{eq:75}
\end{eqnarray}
Here the $\Delta_{mm\sigma}$ ($\Delta_{lm\sigma}$) are the on-site indirect identical orbital (inter-orbital) hybridizations, which relate to the inter-site diagonal (off-diagonal) hopping terms. Under this assumption, there is the relation
\begin{equation}
\sum_kV_{mk\sigma}\ll c_{mk\sigma};f^{\dag}_{m\sigma}\gg = \Delta_{m\sigma}\ll f_{m\sigma};f^{\dag}_{m\sigma}\gg.
\end{equation}

Now let us derive the EOMs for the higher-order Green's functions appearing in Eq.\ \eqref{eq:6} and those newly generated equations, and then implement the decoupling scheme to those three-particle Green's functions to make the equations closed. For the sake of readability, we have put these EOMs in the Appendix.

Solving the derived closed set of equations, we obtain finally the single-particle Green's function for spin $\sigma$ in the $m$-th orbital,
\begin{widetext}
\begin{equation}
\ll f_{m\sigma};f^{\dag}_{m\sigma}\gg=\frac{1+A\big(\bar{n}_{m\sigma'}+I_{1a}\big)+\sum_l \bigg(B\big(\bar{n}_{l\sigma}+I_{1b}\big)+C\big(\bar{n}_{l\sigma'}+I_{1c}\big)\bigg)}
{\omega+\mu-\varepsilon_{fm\sigma}-\Delta_{m\sigma}-A\big(\Delta_{m\sigma}\cdot I_{1a}+I_{2a}\big)-\sum_l\bigg(B\big(\Delta_{m\sigma}\cdot I_{1b}+I_{2b}\big)+C\big(\Delta_{m\sigma}\cdot I_{1c}+I_{2c}\big)\bigg)},\label{eq:singleparticleGF}
\end{equation}
where
\begin{eqnarray}
A&=&U^{eff}_{mm}\bigg/\bigg[\omega+\mu-\varepsilon_{fm\sigma}-U^{eff}_{mm}-\sum_l(U_{lm\sigma\sigma}\bar{n}_{l\sigma}+U_{lm\sigma\sigma'}\bar{n}_{l\sigma'}+2U\bar{n}_{l\sigma}\bar{n}_{l\sigma'})\nonumber\\
&&\qquad\qquad\qquad-\Delta_{m\sigma}-\Delta_{ma\sigma}-\tilde{\Delta}_{ma\sigma}-\sum_l\bar{n}_{m\sigma'}(\Delta_{l\sigma'}+\Delta_{l\sigma})\bigg],\\
B&=&U^{eff}_{lm\sigma\sigma}\bigg/\bigg[\omega+\mu-\varepsilon_{fm\sigma}-U^{eff}_{lm\sigma\sigma}-(U_{mm}\bar{n}_{m\sigma'}+U_{lm\sigma'\sigma}\bar{n}_{l\sigma'}+U\bar{n}_{m\sigma}\bar{n}_{m\sigma'}+U\bar{n}_{l\sigma}\bar{n}_{l\sigma'})\nonumber\\
&&\qquad\qquad\qquad-\sum_{l'}(U_{l'm\sigma\sigma}\bar{n}_{l'\sigma}+U_{l'm\sigma\sigma'}\bar{n}_{l'\sigma'}+2U\bar{n}_{l'\sigma}\bar{n}_{l'\sigma'})-\Delta_{m\sigma}-\Delta_{mb\sigma}-\tilde{\Delta}_{mb\sigma}\nonumber\\
&&\qquad\qquad\qquad-(\bar{n}_{m\sigma'}+\bar{n}_{l\sigma'})\Delta_{m\sigma}-\sum_{l'}(\bar{n}_{l'\sigma}+\bar{n}_{l'\sigma'})\Delta_{m\sigma}\bigg],
\end{eqnarray}
\begin{eqnarray}
C&=&U^{eff}_{lm\sigma'\sigma}\bigg/\bigg[\omega+\mu-\varepsilon_{fm\sigma}-U^{eff}_{lm\sigma'\sigma}-(U_{mm}\bar{n}_{m\sigma'}+U_{lm\sigma\sigma}\bar{n}_{l\sigma}+U\bar{n}_{m\sigma}\bar{n}_{m\sigma'}+U\bar{n}_{l\sigma}\bar{n}_{l\sigma'})\nonumber\\
&&\qquad\qquad\qquad-\sum_{l'}(U_{l'm\sigma\sigma}\bar{n}_{l'\sigma}+U_{l'm\sigma\sigma'}\bar{n}_{l'\sigma'}+2U\bar{n}_{l'\sigma}\bar{n}_{l'\sigma'})-\Delta_{m\sigma}-\Delta_{mc\sigma}-\tilde{\Delta}_{mc\sigma}\nonumber\\
&&\qquad\qquad\qquad-(\bar{n}_{m\sigma'}+\bar{n}_{l\sigma})\Delta_{m\sigma}-\sum_{l'}(\bar{n}_{l'\sigma}+\bar{n}_{l'\sigma'})\Delta_{m\sigma}\bigg],
\end{eqnarray}
\begin{eqnarray}
\Delta_{ma\sigma}&=&\sum_k\frac{{\cal V}^{\ast}_{mmk\sigma'}V_{mk\sigma'}}{\omega+\mu+\varepsilon_{fm\sigma'}-\varepsilon_{fm\sigma}-\varepsilon_{mk\sigma'}}+\sum_{lk,l\neq m}\frac{{\cal V}^{\ast}_{lmk\sigma'}V_{mk\sigma'}}{\omega+\mu+\varepsilon_{fm\sigma'}-\varepsilon_{fm\sigma}-\varepsilon_{mk\sigma'}}\frac{\ll
f_{l\sigma'};f^{\dag}_{m\sigma'}\gg}{\ll
f_{m\sigma'};f^{\dag}_{m\sigma'}\gg},\\
\Delta_{mb\sigma}&=&\sum_k\frac{{\cal V}^{\ast}_{llk\sigma}V_{lk\sigma}}{\omega+\mu+\varepsilon_{fl\sigma}-\varepsilon_{fm\sigma}-\varepsilon_{lk\sigma}}+\sum_{l'k,l'\neq l}\frac{{\cal V}^{\ast}_{l'lk\sigma}V_{lk\sigma}}{\omega+\mu+\varepsilon_{fl\sigma}-\varepsilon_{fm\sigma}-\varepsilon_{lk\sigma}}\frac{\ll
f_{l'\sigma};f^{\dag}_{l\sigma}\gg}{\ll
f_{l\sigma};f^{\dag}_{l\sigma}\gg}, \\
\Delta_{mc\sigma}&=&\sum_k\frac{{\cal V}^{\ast}_{llk\sigma'}V_{lk\sigma'}}{\omega+\mu+\varepsilon_{fl\sigma'}-\varepsilon_{fm\sigma}-\varepsilon_{lk\sigma'}}+\sum_{l'k,l'\neq l}\frac{{\cal V}^{\ast}_{l'lk\sigma'}V_{lk\sigma'}}{\omega+\mu+\varepsilon_{fl\sigma'}-\varepsilon_{fm\sigma}-\varepsilon_{lk\sigma'}}\frac{\ll
f_{l'\sigma'};f^{\dag}_{l\sigma'}\gg}{\ll
f_{l\sigma'};f^{\dag}_{l\sigma'}\gg},\\
\tilde{\Delta}_{ma\sigma}&=&\sum_k\frac{V^{\ast}_{mk\sigma'}{\cal V}_{mmk\sigma'}}{\omega+\mu+\varepsilon_{mk\sigma'}-\varepsilon_{fm\sigma'}-\varepsilon_{fm\sigma}-U_{mma}}+
\nonumber \\
& &
\sum_{kl,l\neq m}\frac{V^{\ast}_{mk\sigma'}{\cal V}_{lmk\sigma'}}{\omega+\mu+\varepsilon_{mk\sigma'}-\varepsilon_{fm\sigma'}-\varepsilon_{fm\sigma}-U_{mma}}\frac{\ll
f_{l\sigma'};f^{\dag}_{m\sigma'}\gg}{\ll
f_{m\sigma'};f^{\dag}_{m\sigma'}\gg},
\end{eqnarray}
\begin{eqnarray}
\tilde{\Delta}_{mb\sigma}&=&\sum_k\frac{V^{\ast}_{lk\sigma}{\cal V}_{llk\sigma}}{\omega+\mu+\varepsilon_{lk\sigma}-\varepsilon_{fl\sigma}-\varepsilon_{fm\sigma}-U_{mmb}}+\sum_{kl',l'\neq l}\frac{V^{\ast}_{lk\sigma}{\cal V}_{l'lk\sigma}}{\omega+\mu+\varepsilon_{lk\sigma}-\varepsilon_{fl\sigma}-\varepsilon_{fm\sigma}-U_{mmb}}\frac{\ll
f_{l'\sigma};f^{\dag}_{l\sigma}\gg}{\ll
f_{l\sigma};f^{\dag}_{l\sigma}\gg},\\ 
\tilde{\Delta}_{mc\sigma}&=&\sum_k\frac{V^{\ast}_{lk\sigma'}{\cal V}_{llk\sigma'}}{\omega+\mu+\varepsilon_{lk\sigma'}-\varepsilon_{fl\sigma'}-\varepsilon_{fm\sigma}-U_{mmc}}+
\nonumber \\ & &
\sum_{kl',l'\neq l}\frac{V^{\ast}_{lk\sigma'}{\cal V}_{l'lk\sigma'}}{\omega+\mu+\varepsilon_{lk\sigma'}-\varepsilon_{fl\sigma'}-\varepsilon_{fm\sigma}-U_{mmc}}\frac{\ll
f_{l'\sigma'};f^{\dag}_{l\sigma'}\gg}{\ll
f_{l\sigma'};f^{\dag}_{l\sigma'}\gg},
\end{eqnarray}
\begin{eqnarray}
I_{1a}&=&\sum_{k}\big(\frac{V^{\ast}_{mk\sigma'}\langle f^{\dag}_{m\sigma'}c_{mk\sigma'}\rangle}{\omega+\mu+\varepsilon_{fm\sigma'}-\varepsilon_{fm\sigma}-\varepsilon_{mk\sigma'}}-\frac{V_{mk\sigma'}\langle c^{\dag}_{mk\sigma'}f_{m\sigma'}\rangle}{\omega+\mu+\varepsilon_{mk\sigma'}-\varepsilon_{fm\sigma'}-\varepsilon_{fm\sigma}-U_{mma}}\big),
\\
I_{2a}&=&\sum_{kk'}\big(-\frac{V^{\ast}_{mk\sigma'}V_{mk\sigma'}\langle c^{\dag}_{mk'\sigma'}c_{mk\sigma'}\rangle}{\omega+\mu+\varepsilon_{fm\sigma'}-\varepsilon_{fm\sigma}-\varepsilon_{mk\sigma'}}
-\frac{V^{\ast}_{mk\sigma'}V_{mk\sigma'}\langle c^{\dag}_{mk\sigma'}c_{mk'\sigma'}\rangle}{\omega+\mu+\varepsilon_{mk\sigma'}-\varepsilon_{fm\sigma'}-\varepsilon_{fm\sigma}-U_{mma}}\big),
\end{eqnarray}
\begin{eqnarray}
I_{1b}&=&\sum_{k}\big(\frac{V^{\ast}_{lk\sigma}\langle f^{\dag}_{l\sigma}c_{lk\sigma}\rangle}{\omega+\mu+\varepsilon_{fl\sigma}-\varepsilon_{fm\sigma}-\varepsilon_{lk\sigma}}-\frac{V_{lk\sigma}\langle c^{\dag}_{lk\sigma}f_{l\sigma}\rangle}{\omega+\mu+\varepsilon_{lk\sigma}-\varepsilon_{fl\sigma}-\varepsilon_{fm\sigma}-U_{mmb}}\big),
\\  
I_{2b}&=&\sum_{kk'}\big(-\frac{V^{\ast}_{lk\sigma}V_{lk\sigma}\langle c^{\dag}_{lk'\sigma}c_{lk\sigma}\rangle}{\omega+\mu+\varepsilon_{fl\sigma}-\varepsilon_{fm\sigma}-\varepsilon_{lk\sigma}}
-\frac{V^{\ast}_{lk\sigma}V_{lk\sigma}\langle c^{\dag}_{lk\sigma}c_{lk'\sigma}\rangle}{\omega+\mu+\varepsilon_{lk\sigma}-\varepsilon_{fm\sigma}-\varepsilon_{fl\sigma}-U_{mmb}}\big),
\\ 
I_{1c}&=&\sum_{k}\big(\frac{V^{\ast}_{lk\sigma'}\langle f^{\dag}_{l\sigma'}c_{lk\sigma'}\rangle}{\omega+\mu+\varepsilon_{fl\sigma'}-\varepsilon_{fm\sigma}-\varepsilon_{lk\sigma'}}-\frac{V_{mk\sigma'}\langle c^{\dag}_{lk\sigma'}f_{l\sigma'}\rangle}{\omega+\mu+\varepsilon_{lk\sigma'}-\varepsilon_{fl\sigma'}-\varepsilon_{fm\sigma}-U_{mmc}}\big),
\\
I_{2c}&=&\sum_{kk'}\big(-\frac{V^{\ast}_{lk\sigma'}V_{lk\sigma'}\langle c^{\dag}_{lk'\sigma'}c_{lk\sigma'}\rangle}{\omega+\mu+\varepsilon_{fl\sigma'}-\varepsilon_{fm\sigma}-\varepsilon_{lk\sigma'}}
-\frac{V^{\ast}_{lk\sigma'}V_{lk\sigma'}\langle c^{\dag}_{lk\sigma'}c_{lk'\sigma'}\rangle}{\omega+\mu+\varepsilon_{lk\sigma'}-\varepsilon_{fl\sigma'}-\varepsilon_{fm\sigma}-U_{mmc}}\big),
\end{eqnarray}
where
\begin{eqnarray}
U_{mma}&=&U^{eff}_{mm}+2\sum_l(U_{lm\sigma\sigma}\bar{n}_{l\sigma}+U_{lm\sigma\sigma'}\bar{n}_{l\sigma'}+2U\bar{n}_{l\sigma}\bar{n}_{l\sigma'}),\\
U_{mmb}&=&U^{eff}_{lm\sigma\sigma}+2(U_{mm}\bar{n}_{m\sigma'}+U_{lm\sigma\sigma'}\bar{n}_{l\sigma'}+U\bar{n}_{m\sigma}\bar{n}_{m\sigma'}+U\bar{n}_{l\sigma}\bar{n}_{l\sigma'})\nonumber\\
&&+2\sum_{l'}(U_{l'm\sigma\sigma}\bar{n}_{l'\sigma}+U_{l'm\sigma\sigma'}\bar{n}_{l'\sigma'}+2U\bar{n}_{l'\sigma}\bar{n}_{l'\sigma'}),\\
U_{mmc}&=&U^{eff}_{lm\sigma\sigma'}+2(U_{mm}\bar{n}_{m\sigma'}+U_{lm\sigma\sigma}\bar{n}_{l\sigma}+U\bar{n}_{m\sigma}\bar{n}_{m\sigma'}+U\bar{n}_{l\sigma}\bar{n}_{l\sigma'})\nonumber\\
&&+2\sum_{l'}(U_{l'm\sigma\sigma}\bar{n}_{l'\sigma}+U_{l'm\sigma\sigma'}\bar{n}_{l'\sigma'}+2U\bar{n}_{l'\sigma}\bar{n}_{l'\sigma'}).
\end{eqnarray}
%
In the above equations, $l\neq l'$. Moreover, in the numerical calculations we will use the following Hermitian relations
\begin{eqnarray}
\langle f^{\dag}_{m\sigma'}c_{mk\sigma'}\rangle=\langle c^{\dag}_{mk\sigma'}f_{m\sigma'}\rangle,\qquad \langle c^{\dag}_{mk'\sigma'}c_{mk\sigma'}\rangle=\langle c^{\dag}_{mk\sigma'}c_{mk'\sigma'}\rangle.\nonumber
\end{eqnarray}
\end{widetext}


As we have mentioned in Ref.\ \onlinecite{FO11}, the hybridization functions in Eq.\ \eqref{eq:singleparticleGF} will be obtained along with the DMFT self-consistency conditions.
For the Bethe lattice, when one neglects the inter-site hoppings between different orbitals,
the DMFT self-consistency condition should be
\begin{eqnarray}
\Delta_{m\sigma}=t^2_m \ll f_{m\sigma};f^{\dag}_{m\sigma}\gg.
\end{eqnarray}
When we take now into account the inter-site hoppings between different orbitals,
we obtain that the DMFT self-consistency condition will be
\begin{eqnarray}
\Delta_{m\sigma} &=& t_m^2\frac{t^2_m}{t^2_{tot}}\ll f_{m\sigma};f^{\dag}_{m\sigma}\gg\nonumber\\
&+&\sum_{l,l\neq m} t^2_m \frac{t^2_l}{t^2_{tot}}\ll f_{l\sigma};f^{\dag}_{l\sigma}\gg,~~\label{eq:dmftselfconsistencycondition}
\end{eqnarray}
with $t_{tot}$ is the total amplitude summing over all the orbitals, i.e.,
 $t_{tot}=\sum_{m}t_m$. All the other hybridization functions can be obtained by interpolation.

In this work, besides the inter-site inter-orbital hoppings, we concentrate and fully include the inter-orbital fluctuations, because the Coulomb interactions are the most basic interactions and give the principal contributions, and usually are also the only interactions incorporated in most LDA+DMFT works.\cite{Kunes,Pavarini,yzhang,Held} Other interactions, such as e.g. the spin-flip term and pair-hopping term\cite{CFCT09}, are considered to contribute less and contribute mainly to the shape of the Hubbard bands (for a detailed explanation, see Ref.\ \onlinecite{FO11}).


\section{Results and discussions\label{sect3}}
\begin{figure}
\includegraphics[width=6cm,angle=-90]{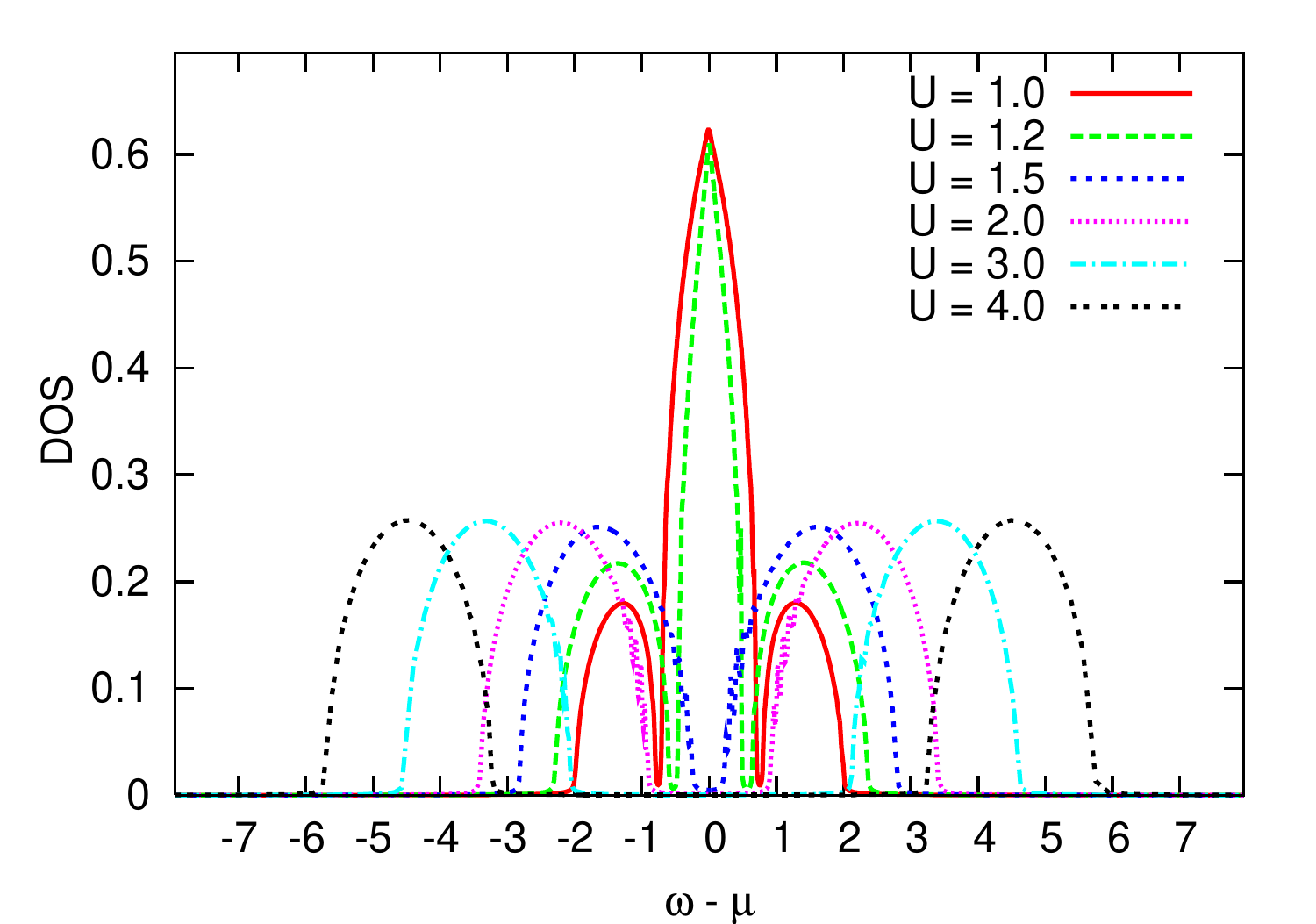}%
\caption{(Color online) Quasiparticle densities of states for the halffilled two-orbital Hubbard model on the Bethe lattice, computed with the MO-EOM method for the parameters: $D_1=D_2=1$, $T=0.01$, and $J=0$.\label{fig1}}
\end{figure}

To investigate the accuracy of the derived  method, we have studied
the two-orbital Hubbard model  in the halffilled, paramagnetic case with this MO-EOM impurity solver. The DOS for the quasiparticles are shown in  Fig.~\ref{fig1}, computed for the case where the band widths of the two orbitals are identical and the Hund's coupling constant is zero. It is observed that, along with the increase of the Coulomb interaction strength $U$, the system turns from the metallic state to the insulating state. The Mott metal-insulator transition occurs at nearly $U=1.45$. We mention that the numerical calculation automatically fulfills the particle-hole symmetry and the sum rule that the integral over all the DOS equals identity.

\begin{figure}
\includegraphics[width=6cm,angle=-90]{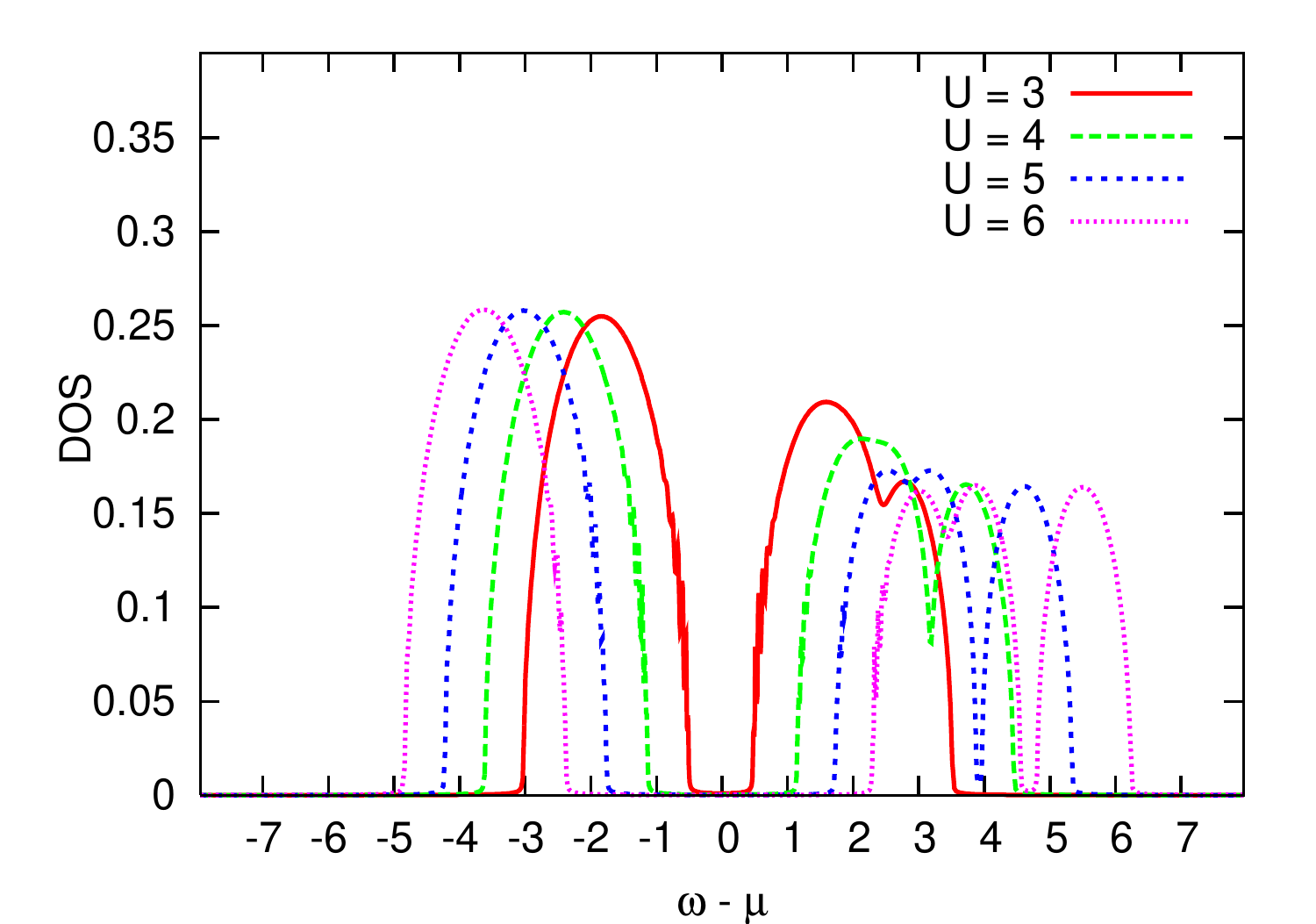}%
\caption{(Color online) Densities of states calculated for the halffilled two-orbital Hubbard model on the Bethe lattice for different $U$, with the parameters: $D_1=D_2=1$, $T=0.01$, and $J=U/4$.\label{fig2}}
\end{figure}

In Fig.\ \ref{fig2} we show again the densities of states of the quasiparticles for the two-orbital Hubbard model at the halffilling. Now we have chosen the parameters such that the two orbitals have identical band widths, i.e., $D_1=D_2=1$, and the Hund's coupling constant $J=U/4$. The results for larger $U$ reveal that the upper Hubbard band is in fact composed of three sub-peaks which are generated correspondingly by the intra-orbital and two types of inter-orbital Coulomb interactions. With the increase of $U$, these three sub-peaks are split more, due to the difference between the intra-orbital and inter-orbital Coulomb interaction strengths, i.e., the existence of the nonzero Hund's coupling constant.
The result in Fig.~\ref{fig2} is obtained from the evolution with Eq.\ \eqref{eq:singleparticleGF}, and it follows the sum rule.
The word `evolution' is here used for the iterative approach, iterating from an initial guessed Green's function to a final converged result. Fig.\ \ref{fig2} shows however an asymmetric DOS for the lower and upper Hubbard bands, something which
is unusual and requires further investigations. As outlined in the following, this is related to how the iterative algorithm is performed.

\begin{figure}
\includegraphics[width=6cm,angle=-90]{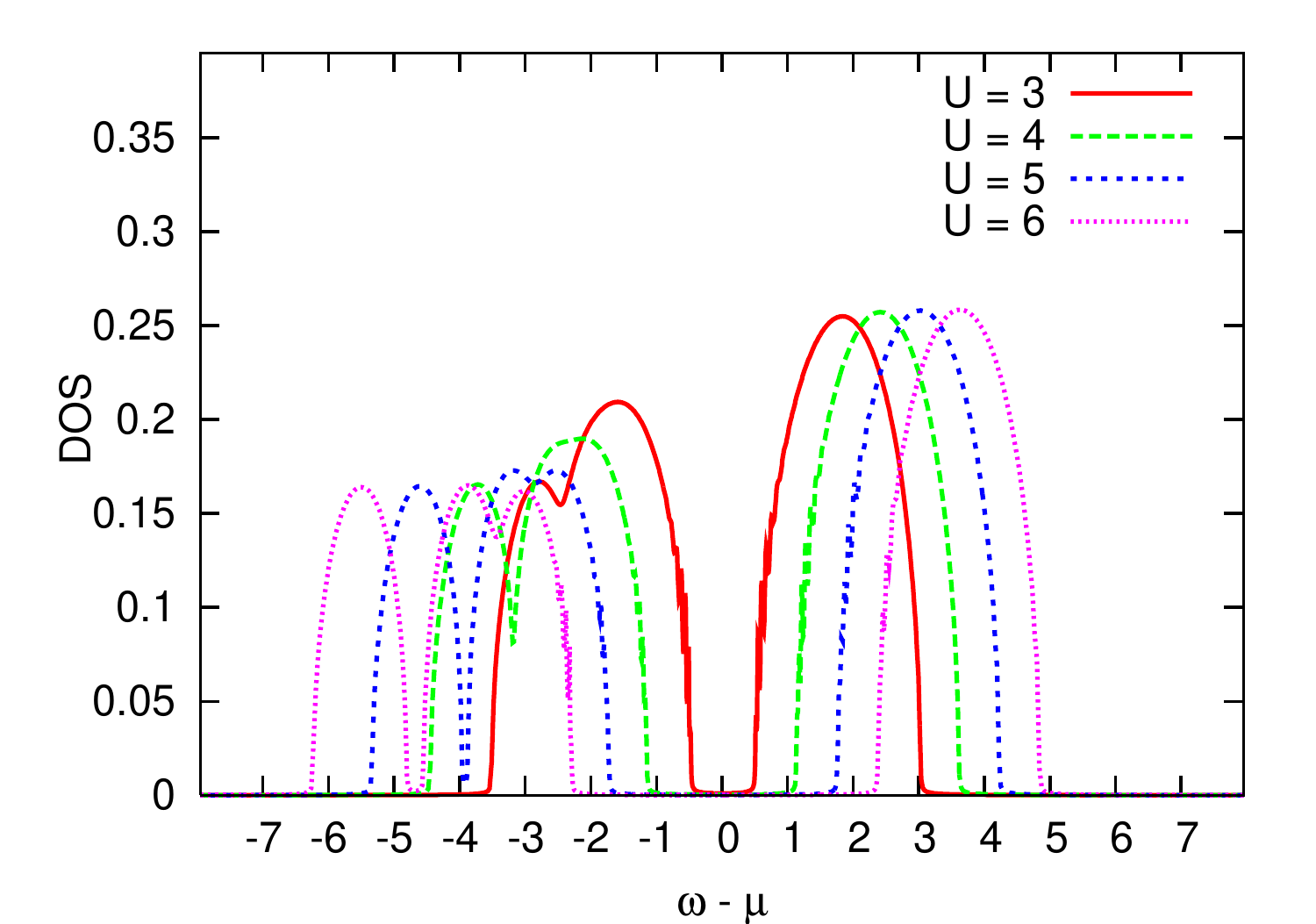}%
\caption{(Color online) Computed densities of states for the halffilled two-orbital `hole' system on the Bethe lattice for different $U_{hole}$, obtained with the parameters: $D_1=D_2=1$, $T=0.01$, and $J_{hole}=U_{hole}/4=-U/4$.\label{fig3}}
\end{figure}
\begin{figure}
\includegraphics[width=6cm,angle=-90]{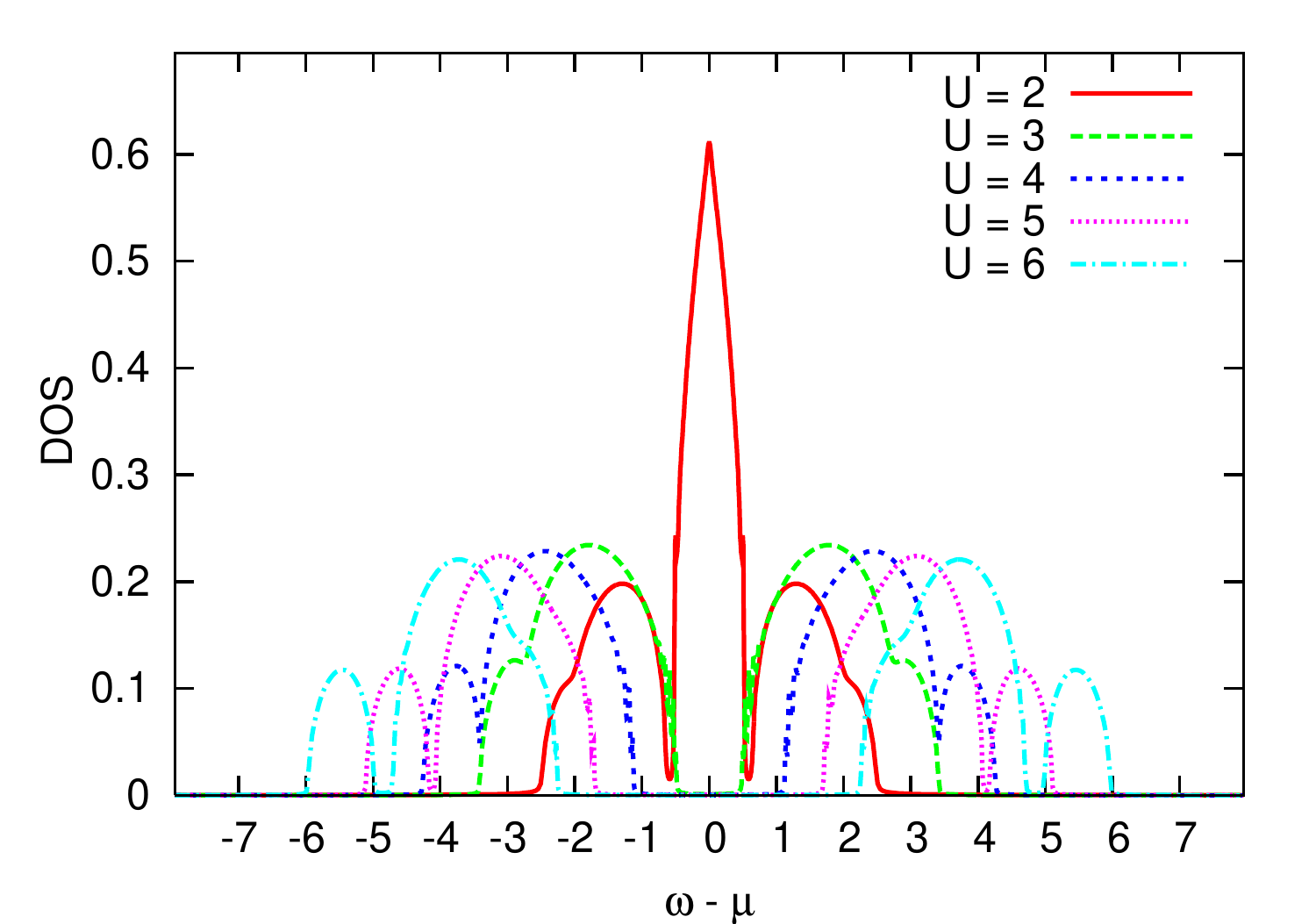}%
\caption{(Color online) Quasiparticle densities of states for the halffilled two-orbital Hubbard model on the Bethe lattice with particle-hole symmetry at different $U$, with the parameters: $D_1=D_2=1$, $T=0.01$, and $J=U/4$.\label{fig4}}
\end{figure}

To study in more detail the issue of the particle-hole symmetry, we considering an electron system where all the states are initially fully occupied, the presence of a `hole' state (empty of electron) is equivalently a fermionic quasiparticle with spin index. Therefore, a many-electron system can be equivalently seen as a many-hole system, where the equivalent `hole' system is similar to the electron system but only the interactions between the `hole' states are with negative Coulomb interaction strengths. All our derived and obtained equations are also valid for these `hole' states. Thus, we can implement our equations of motion to this `hole' system and set the hole position for $\sigma$ spin channel in the $m$-th orbital as $-\varepsilon_{mf\sigma}$. When $J_{hole}=U_{hole}/4=-U/4$, we obtain consequently the DOS of the `hole' quasiparticles, as shown in Fig.~\ref{fig3}.
However, the studied `hole' system and the electron system are actually the same system, but are seen from different view angles. Therefore, the DOS obtained from the `hole' system is also the DOS of the electrons. For any system, when the final self-consistently converged Green's function is obtained, no matter whether we calculate one next iteration as an electron system or as a `hole' system, it will give identical results, i.e., it will follow the particle-hole symmetry.
One should note that this reasoning is not only applicable for the halffilled case, but for all arbitrary occupations, i.e.,
all the systems should follow this rule. Combining this idea into the evolution, we have calculated both the electron system and the `hole' system in our genetic algorithm scheme,\cite{FZJ09} which can also be considered to be a double-check procedure, and obtained the corrected DOS, shown in Fig.\ \ref{fig4}, for the halffilled two-orbital system with nonzero $J=U/4$. The particle-hole symmetry is well recovered. We mentioned that both the lower and upper Hubbard bands show a splitting in sub-peaks for a large $U$ due to the nonzero $J$.
The lower Hubbard band is  split, whereas in the DMFT self-consistency condition Eq.\ \eqref{eq:dmftselfconsistencycondition} for the Bethe lattice, the updated hybridization functions can not offer the information of this splitting. Combining the calculation of `hole' states into the evolution algorithm is a double-check process which can indeed safeguard the accuracy of the finally obtained result. In the genetic algorithm,\cite{GA1} this operation has the meaning to increase the diversity of the trial Green's functions, something which is important for the genetic algorithm so that the system will iteratively evolve in the right direction. If employed with linear mixing, the usual form of a searching scheme is
\begin{eqnarray}
G^{n+1}_f=\alpha G^{new}_f+(1-\alpha)G^n_f,
\end{eqnarray}
which means that in each iteration only a small amount of newly generated Green's function will be mixed with the Green's function of the last iteration so that the integral equations are iteratively solved. $\alpha$ (and also the $\beta$ below) is the mixing parameter and $n$ is the iteration number.
Our discourse above purports that now it should be modified to
\begin{eqnarray}
G^{n+1}_f=\alpha G^{new}_f+\beta G^{new}_{hole}+(1-\alpha-\beta)G^n_f.
\end{eqnarray}
In the genetic algorithm this does not cost additional time because in each iteration we will calculate a group of Green's functions. We can simply change some calculations of electron system to `hole' system. For details of the genetic algorithm for a continuous spectrum, see Ref.\ \onlinecite{phdthesis-feng}. If the iterative search is performed with linear mixing,  the required time will be doubled in serial calculation, but it is still much faster than other numerical methods.


\begin{figure}
\includegraphics[width=6cm,angle=-90]{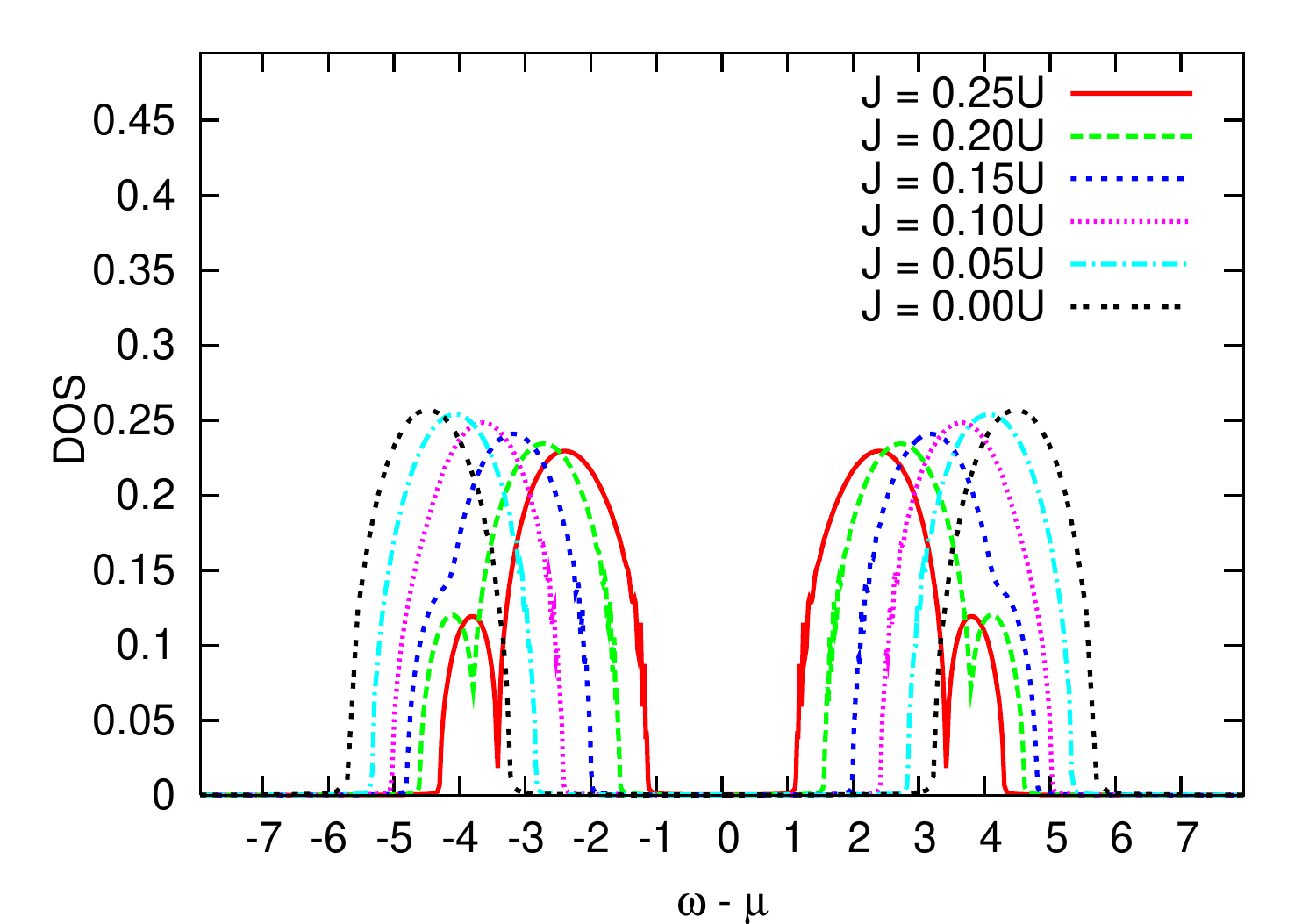}%
\caption{(Color online) Quasiparticle densities of states computed
for the halffilled two-orbital Hubbard model on the Bethe lattice at different $J$, for the parameters: $D_1=D_2=1$, $T=0.01$, and $U=4$. \label{fig5}}
\end{figure}
Fig.\ \ref{fig5} shows the quasiparticle DOS at a fixed $U$ but with different $J$, where the influence of the choice of $J$ is studied. It can be observed that both the lower and upper Hubbard bands are split into sub-peaks by the Coulomb interactions and their splitting increases  with the increase of the Hund's coupling constant $J$. Here we have chosen $U=4$ which is a medium value. Taken the information in this figure together with that in Fig.\ \ref{fig4}, it is apparent that for a small $U$ the Hubbard bands can not be clearly split and will acquire only  a change of the overall shape. For very large $U$, however, both the lower and upper Hubbard band can become split even into three peaks.

\vspace{3mm}
\begin{figure}
\includegraphics[width=6cm,angle=-90]{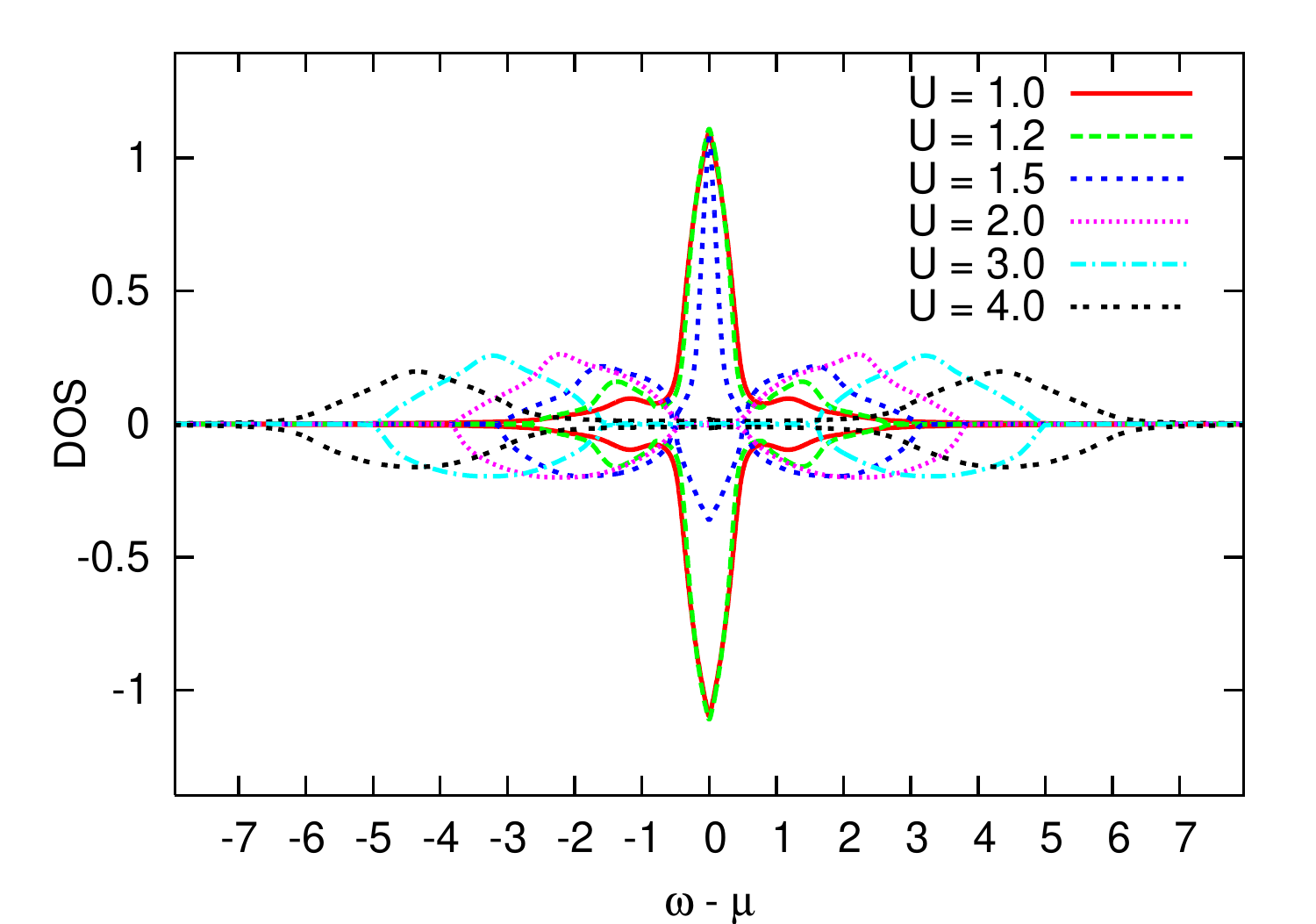}%
\caption{(Color online) Densities of states for the halffilled two-orbital Hubbard model on the Bethe lattice at different $U$, computed with the MO-EOM approach for two different band-widths, using  the parameters: $D_2=2D_1=2$, $T=0.01$, and $J=0$. The DOS shown on the negative ordinate are those obtained for the wide orbital, those on the positive ordinate are obtained for the narrow orbital.\label{fig6}}
\end{figure}
In the above calculations, we studied the situation that the two orbitals have identical band-widths. Now let us turn to the case that the band-widths of the two orbitals are different, which can be occur, for example, for those systems that have both partial filled $d$ and $f$ electron shells, because the $f$ electrons are considered to be more localized and should thus have a narrower band-width. To study this case we have used completely the same code, but only changed the parameters. Once a method can be generalized to more systems but without any additional work needed, it will be much easier to work with and will encounter more applications in LDA+DMFT calculations for real materials. In Fig.\ \ref{fig6}, we present the quasiparticle DOS for both the narrow and wide orbital, with the parameters such that the band-width of the wide orbital is twice that of the narrow orbital, and the Coulomb interaction strengths are identical, i.e., $J=0$. For sake of clarity we plotted the DOS of the narrow orbital on positive $y$ axis and that of the wide orbital on the negative $y$ axis. For the same $U$ value, the narrow and the wide orbitals have different DOS. Also, we studied the influence of $U$. With an increase of $U$, the narrow and wide orbitals simultaneously change both from being metallic states to insulating states. Hence, no orbital-selective Mott transition is observed.

\begin{figure}
\includegraphics[width=6cm,angle=-90]{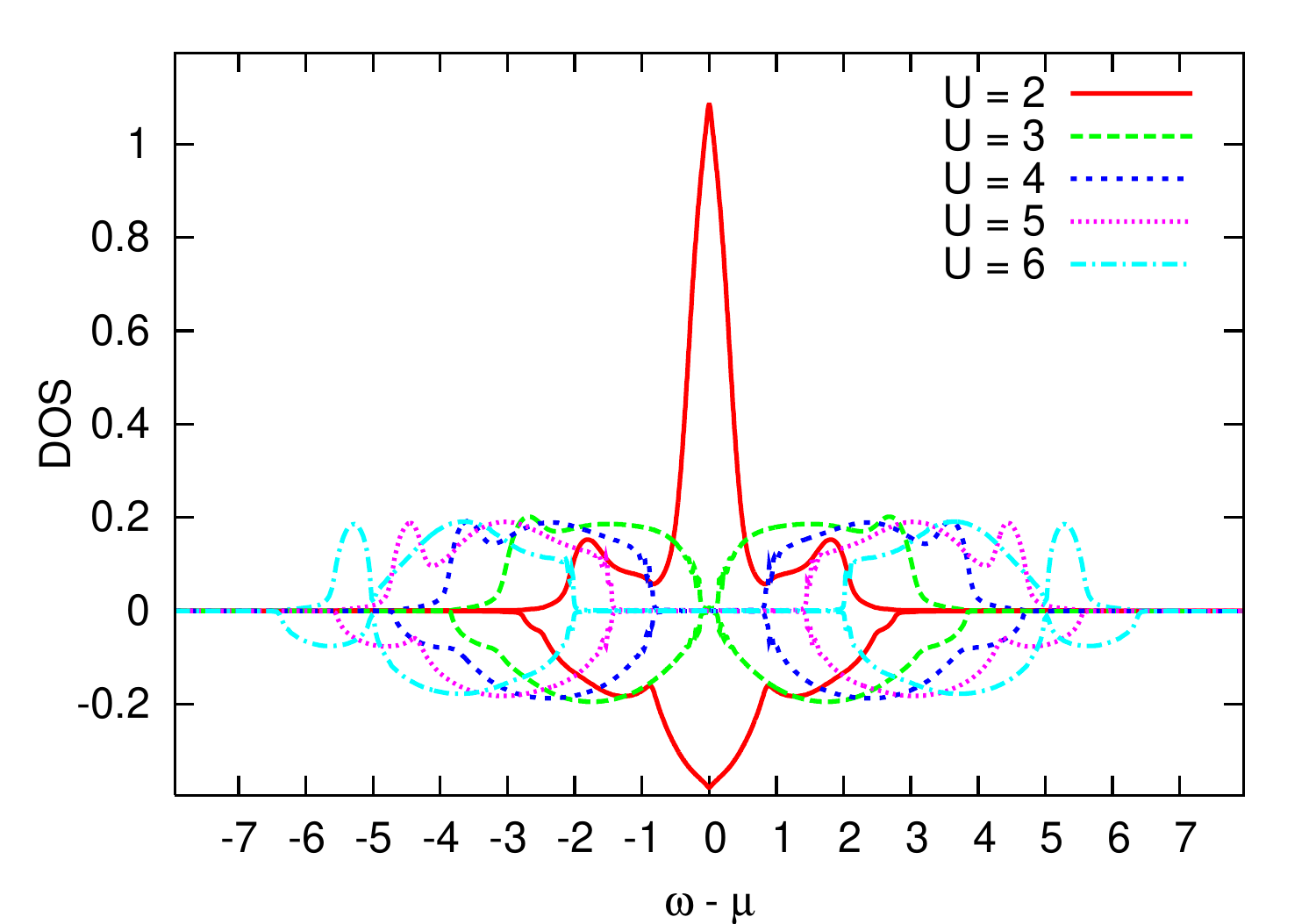}%
\caption{(Color online) Densities of states for the halffilled two-orbital Hubbard model on the Bethe lattice at different $U$, computed with the parameters: $D_2=2D_1=2$, $T=0.01$, and $J=U/4$. The DOS shown on the negative ordinate is that for the wide orbital, that on the positive ordinate for the narrow orbital.\label{fig7}}
\end{figure}
Fig.\ \ref{fig7} shows the quasiparticle DOS for the paramagnetic two-orbital Hubbard model with different band-widths for the two orbitals, now computed with the parameter $J=U/4$. Similar features as shown in Fig.\ \ref{fig6} can be observed, but at a higher critical value of $U$ compared to the results in Fig.\ \ref{fig6}, which is due to the decrease of the effective Coulomb interaction strength caused by the nonzero $J$. In addition,  the Hubbard bands display a splitting into sub-peaks in the large $U$ region.

\begin{figure}
\includegraphics[width=6cm,angle=-90]{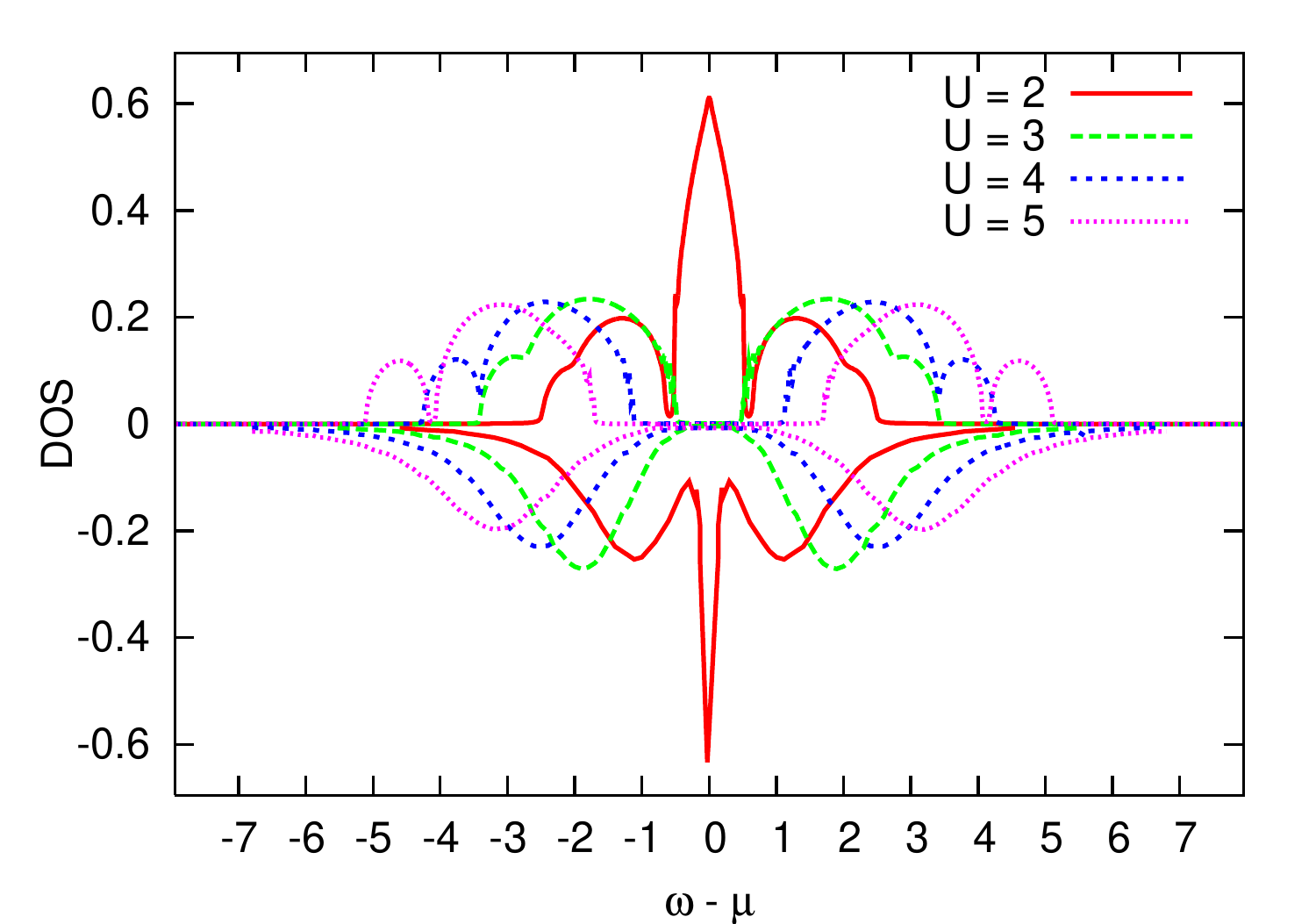}%
\caption{(Color online) Comparison of the densities of states for the halffilled two-orbital Hubbard model on the Bethe lattice calculated with the MO-EOM to those computed with the NRG method for various $U$, using the parameters: $D_1=D_2=1$, $T=0.01$, and $J=U/4$. The DOS shown on the negative ordinate are those of the NRG method, those on the positive ordinate are computed with our EOM method. The NRG data are extracted from Ref.\ \onlinecite{PruschkeNRG}. \label{fig8}}
\end{figure}
Next, in order to address the accuracy of our MO-EOM impurity solver, we give some comparisons of our numerical results to those obtained with other methods. First, we present a comparison to the numerical renormalization group method in Fig.\ \ref{fig8} for the paramagnetic two-orbital system where the two orbitals have identical band-widths, i.e., $D_1=D_2=1$. The NRG data are extracted from Ref.\ \onlinecite{PruschkeNRG}. For the purpose of visibility, the NRG results are plotted on the negative $y$ axis, while those of our MO-EOM method are plotted on the positive $y$ axis. The quasiparticle DOS obtained with the same $U$ are plotted in the same color. Overall, our MO-EOM method and the NRG method show a very good agreement no matter if it is on the widths and positions of the Hubbard bands or on the critical value of $U$ for the Mott metal-insulator transition. In addition, our MO-EOM result displays not only more micro-structures of the DOS caused by the nonzero $J$ of the anistropic Coulomb interactions, but also shows a greatly reduced tail effect, which is an advantage of our EOM method that in our genetic algorithm evolution scheme the Lorentzian broadening can even be set  as zero. With regard to these two points, our MO-EOM method has a computational advantage and is  more accurate.

\begin{figure}[tbh]
\includegraphics[width=10cm,angle=-90]{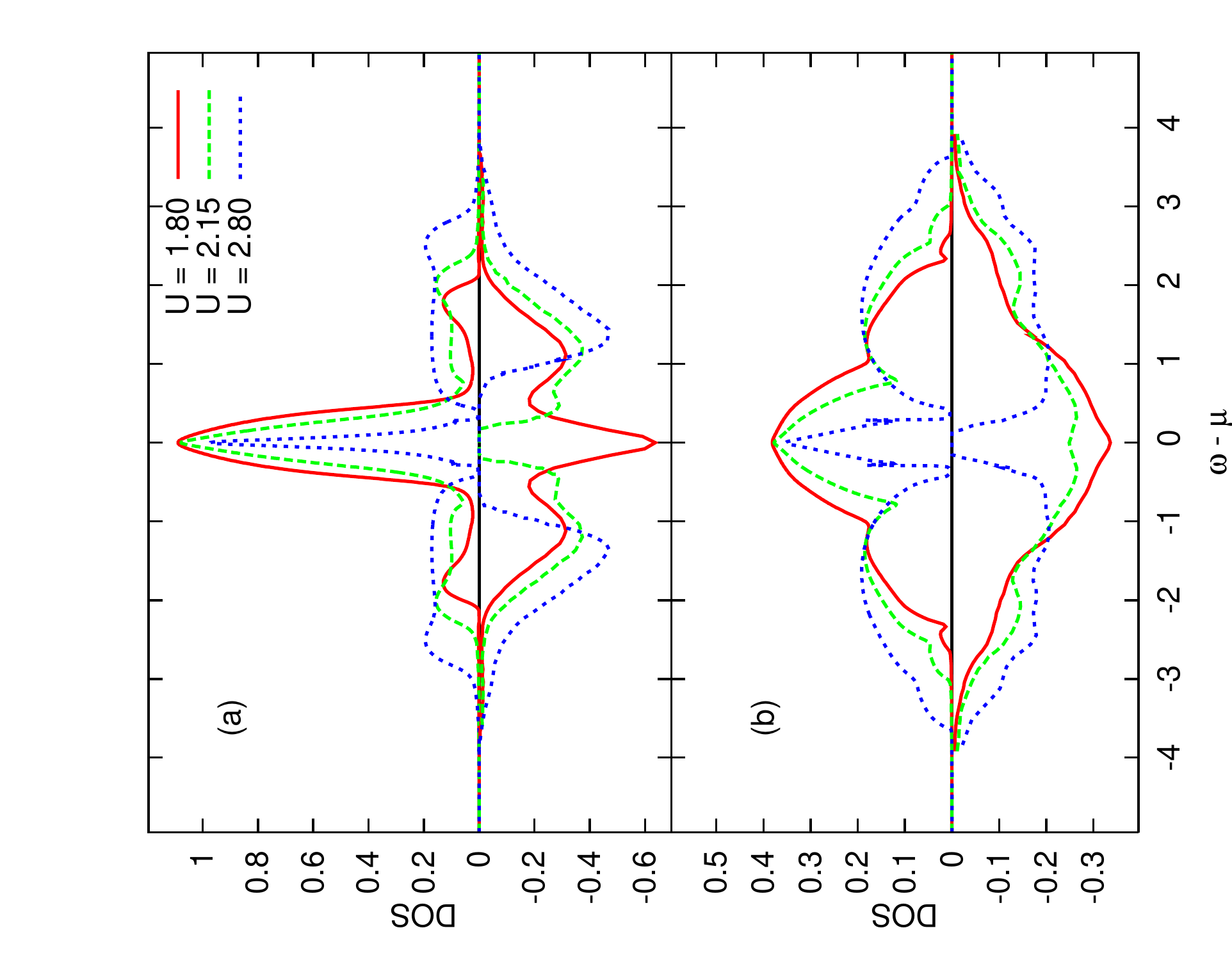}%
\caption{(Color online) Comparison of the densities of states obtained with the MO-EOM and QMC methods for the halffilled two-orbital Hubbard model on the Bethe lattice for different $U$, with the parameters: $D_2=2D_1=2$, $T=1/40$, and $J=U/4$. In panel (a), the quasiparticle
DOS are compared between our MO-EOM method and the QMC method for the narrow orbital. In panel (b), the DOS are compared between our method and the QMC method for the wide orbital. In both panels, the DOS shown on the negative ordinate are from the QMC method and those on the positive ordinate are from our MO-EOM method. The QMC data are those from Ref.\ \onlinecite{OSMT1}.\label{fig9}}
\end{figure}

Next, we provide a comparison of the MO-EOM quasiparticle DOS to values obtained with the QMC method in Fig.\ \ref{fig9} for the case that $J=U/4$.
The QMC data are obtained from Ref.\ \onlinecite{OSMT1}, and all parameters are taken to be  the same. In panel (a), we compare the DOS of the narrow orbital correspondingly obtained with our MO-EOM method and the QMC method, where the QMC data are plotted on the negative $y$ axis and the MO-EOM data on the positive $y$ axis. In panel (b), the DOS of the wide orbital are compared between the QMC and our MO-EOM method, and similarly here the QMC and MO-EOM data are plotted on negative and positive $y$ axis, respectively. To some extent, both methods have an agreement on the Hubbard bands, for the positions of the orbital levels. However, the two methods show a different critical value of $U$ and different behavior when the system approaches the metal-insulator transition, which has been found in  a previous study, \cite{FO11} too,  and is here confirmed.
The reason for the different behavior is that, in our MO-EOM method, we have taken into account the inter-site inter-orbital hopping effects (which are also mentioned in Ref.~\onlinecite{Held}),
and the DOS obtained with the two methods are actually calculated with a different Hamiltonian\cite{PruschkeNRG,OSMT1,Liebsch}(see Ref.~\onlinecite{FO11}). 
The inter-site hopping reflects the interactions between the orbitals at the studied site and the orbitals of other sites. But, when all the sites are identical, due to the translation invariance, the inter-site inter-orbital hopping  can equivalently be seen as an interaction between one orbital and all other on-site orbitals, whose influence will suppress the orbital-selective Mott transition. Moreover, we have taken into account the inter-orbital fluctuations. For spin $\sigma$ in the $m$-th orbital, the spins existing in other orbitals have a  similar contribution as spin $\sigma'$ in the $m$-th orbital, which can be seen from our equations. In Fig.\ \ref{fig2} we have noted that, for the two-orbital paramagnetic system, the Hubbard band for one spin is actually composed of three parts corresponding to the other three spins. Accordingly, the Kondo peak should also be composed of three parts. This means that, once there is a Kondo peak for one spin, there will be a Kondo peak for another spin. Therefore, the existence of the on-site inter-orbital fluctuations will suppress the orbital-selective Mott transition, too.  We also note that in our MO-EOM result the Kondo peak is nearly pinned at the Fermi level, which is an important aspect for an impurity solver because in theory the Kondo peak should be pinned at one point at the Fermi level,\cite{Kondopeak} while the QMC method does not show this feature.

\section{Summary\label{sect4}}
In this paper we have proposed an improved fast and applicable multi-orbital impurity solver for the DMFT based on the equations of motion method with fully including the inter-orbital fluctuations beyond the mean-field approximation treatment to the inter-orbital Coulomb interactions.
Our derivation has provided an alternative, fast and more precise way to obtain the local single-particle Green's functions.

The derived impurity solver works directly on the real frequencies axis and for any temperatures.
Comparing with other methods, it is less memory and cpu time expensive.
It has also given improved detail in the micro-structures of the DOS and shows an improved performance in reducing the tail effect.
Furthermore, it still keeps the generality of the parameters. From our investigations
it appears that our MO-EOM method can be generally applied to a large range of multi-orbital systems,
no matter if these have identical  or different band-widths, nonzero $J$ or not. In the present work
 we only studied the paramagnetic case, but the MO-EOM method
can also be applied for the magnetic case. The comparisons of our numerical results with those obtained with NRG and QMC methods attest that our method shows not only a good agreement but also more precise micro-structures of the Hubbard bands, and also some distinct different behavior in approaching the Mott metal-insulator transition with the self-consistent inclusion of the inter-site inter-orbital hoppings and the on-site inter-orbital Coulomb interactions. Therefore, our MO-EOM method appears to be a good candidate for a fast impurity solver for the dynamical mean field theory
and for LDA+DMFT calculations of multi-orbital strongly correlated electron systems.\cite{code}

\acknowledgments
{
Q.\ Feng would like to thank H.\ O.\ Jeschke and Y.\ -Z.\ Zhang for useful discussions at an early stage of this work, performed as a part of Q.~Feng's PhD study at the University Frankfurt, and was supported by the DFG Emmy Noether program under Project ``JE~277/1-2".
This work is supported by the Swedish Research Council (VR) and SKB.
The calculations have been supported through computer time from the Swedish National Infrastructure for Computing (SNIC)
and have been performed at the Swedish national computer centers NSC, UPPMAX, and HPC2N.}

\appendix
\section{Equations of motion for the two-particle Green's functions}

In order to make the paper more easily readable, we present here the equations of motion for the higher-order two-particle Green's functions that are needed to derive the final single-particle Green's function Eq.\ (\ref{eq:singleparticleGF}) given in this paper. In addition we also list the decoupling scheme for the three-particle Green's functions appearing in those equations of motion.

The EOMs of the higher-order Green's functions are
\begin{widetext}
\begin{eqnarray}
(\omega+\mu-\varepsilon_{fm\sigma})G^{m}_{nf}
&=&\bar{n}_{m\sigma'}+U_{mm}\ll~\hat{n}_{m\sigma'}f_{m\sigma};f^{\dag}_{m\sigma}\gg\nonumber\\
&+&\sum_l\big(U_{lm\sigma\sigma}\ll~\hat{n}_{l\sigma}\hat{n}_{m\sigma'}f_{m\sigma};f^{\dag}_{m\sigma}\gg+U_{lm\sigma'\sigma}\ll
\hat{n}_{l\sigma'}\hat{n}_{m\sigma'}f_{m\sigma};f^{\dag}_{m\sigma}\gg\big)\nonumber\\
&+&\sum_k\big(-V^{\ast}_{mk\sigma'}\ll
c^{\dag}_{mk\sigma'}f_{m\sigma'}f_{m\sigma};f^{\dag}_{m\sigma}\gg+V_{mk\sigma}\ll
\hat{n}_{m\sigma'}c_{mk\sigma};f^{\dag}_{m\sigma}\gg\nonumber\\
&+&V_{mk\sigma'}\ll
f^{\dag}_{m\sigma'}c_{mk\sigma'}f_{m\sigma};f^{\dag}_{m\sigma}\gg\big)\label{eq:9}
\\
(\omega+\mu-\varepsilon_{fm\sigma})G^{lm}_{n\sigma f}
&=&\bar{n}_{l\sigma}+U_{lm\sigma\sigma}\ll~\hat{n}_{l\sigma}f_{m\sigma};f^{\dag}_{m\sigma}\gg\nonumber\\
&+&\big(U_{mm}\ll~\hat{n}_{m\sigma'}\hat{n}_{l\sigma}f_{m\sigma};f^{\dag}_{m\sigma}\gg+U_{lm\sigma'\sigma}\ll
\hat{n}_{l\sigma'}\hat{n}_{l\sigma}f_{m\sigma};f^{\dag}_{m\sigma}\gg\big)\nonumber\\
&+&\sum_{\substack{l'\\l'\neq l, l'\neq m}}\big(U_{l'm\sigma\sigma}\ll~\hat{n}_{l\sigma}\hat{n}_{l\sigma}f_{m\sigma};f^{\dag}_{m\sigma}\gg+U_{l'm\sigma'\sigma}\ll
\hat{n}_{l'\sigma'}\hat{n}_{l\sigma}f_{m\sigma};f^{\dag}_{m\sigma}\gg\big)\nonumber\\
&+&\sum_k\big(-V^{\ast}_{lk\sigma}\ll
c^{\dag}_{lk\sigma}f_{l\sigma}f_{m\sigma};f^{\dag}_{m\sigma}\gg+V_{mk\sigma}\ll
\hat{n}_{m\sigma'}c_{mk\sigma};f^{\dag}_{m\sigma}\gg\nonumber\\
&+&V_{lk\sigma}\ll
f^{\dag}_{l\sigma}c_{lk\sigma}f_{m\sigma};f^{\dag}_{m\sigma}\gg\big)\label{eq:9a}
\end{eqnarray}
\begin{eqnarray}
(\omega+\mu-\varepsilon_{fm\sigma})G^{lm}_{n\sigma' f}
&=&\bar{n}_{l\sigma'}+U_{lm\sigma'\sigma}\ll~\hat{n}_{l\sigma'}f_{m\sigma};f^{\dag}_{m\sigma}\gg\nonumber\\
&+&\big(U_{mm}\ll~\hat{n}_{m\sigma'}\hat{n}_{l\sigma'}f_{m\sigma};f^{\dag}_{m\sigma}\gg+U_{lm\sigma\sigma}\ll
\hat{n}_{l\sigma}\hat{n}_{l\sigma'}f_{m\sigma};f^{\dag}_{m\sigma}\gg\big)\nonumber\\
&+&\sum_{\substack{l'\\l'\neq l, l'\neq m}}\big(U_{l'm\sigma\sigma}\ll~\hat{n}_{l\sigma}\hat{n}_{l\sigma'}f_{m\sigma};f^{\dag}_{m\sigma}\gg+U_{l'm\sigma'\sigma}\ll
\hat{n}_{l'\sigma'}\hat{n}_{l\sigma'}f_{m\sigma};f^{\dag}_{m\sigma}\gg\big)\nonumber\\
&+&\sum_k\big(-V^{\ast}_{lk\sigma'}\ll
c^{\dag}_{lk\sigma'}f_{l\sigma'}f_{m\sigma};f^{\dag}_{m\sigma}\gg+V_{mk\sigma}\ll
\hat{n}_{m\sigma'}c_{mk\sigma};f^{\dag}_{m\sigma}\gg\nonumber\\
&+&V_{lk\sigma'}\ll
f^{\dag}_{l\sigma'}c_{lk\sigma'}f_{m\sigma};f^{\dag}_{m\sigma}\gg\big)\label{eq:9b}
\end{eqnarray}
\begin{eqnarray}
(\omega+\mu-\varepsilon_{mk\sigma})G^{m}_{nc}
&=&V^{\ast}_{mmk\sigma}\ll~\hat{n}_{m\sigma'}f_{m\sigma};f^{\dag}_{m\sigma}\gg
+\sum_{k'}\big(V^{\ast}_{mk'\sigma'}\ll~f^{\dag}_{m\sigma'}c_{mk'\sigma'}c_{mk\sigma};f^{\dag}_{m\sigma}\gg\nonumber\\
&-&V_{mk'\sigma'}\ll~c^{\dag}_{mk'\sigma'}f_{m\sigma'}c_{mk\sigma};f^{\dag}_{m\sigma}\gg\big)
+\sum_{l,l\neq m} V^{\ast}_{lmk\sigma}\ll~\hat{n}_{m\sigma'}f_{l\sigma};f^{\dag}_{m\sigma}\gg\label{eq:10}
\end{eqnarray}
\begin{eqnarray}
(\omega+\mu-\varepsilon_{mk\sigma})G^{lm}_{n\sigma c}
&=&V^{\ast}_{mmk\sigma}\ll~\hat{n}_{l\sigma}f_{m\sigma};f^{\dag}_{m\sigma}\gg
+\sum_{k'}\big(V^{\ast}_{lk'\sigma}\ll~f^{\dag}_{l\sigma}c_{lk'\sigma}c_{mk\sigma};f^{\dag}_{m\sigma}\gg\nonumber\\
&-&V_{lk'\sigma}\ll~c^{\dag}_{lk'\sigma}f_{l\sigma}c_{mk\sigma};f^{\dag}_{m\sigma}\gg\big)
+\sum_{l',l'\neq m} V^{\ast}_{l'mk\sigma}\ll~\hat{n}_{l\sigma}f_{l'\sigma};f^{\dag}_{m\sigma}\gg\label{eq:10a}
\end{eqnarray}
\begin{eqnarray}
(\omega+\mu-\varepsilon_{mk\sigma})G^{lm}_{n\sigma' c}
&=&V^{\ast}_{mmk\sigma}\ll~\hat{n}_{l\sigma'}f_{m\sigma};f^{\dag}_{m\sigma}\gg
+\sum_{k'}\big(V^{\ast}_{lk'\sigma'}\ll~f^{\dag}_{l\sigma'}c_{lk'\sigma'}c_{mk\sigma};f^{\dag}_{m\sigma}\gg\nonumber\\
&-&V_{lk'\sigma'}\ll~c^{\dag}_{lk'\sigma'}f_{l\sigma'}c_{mk\sigma};f^{\dag}_{m\sigma}\gg\big)
+\sum_{l',l'\neq m} V^{\ast}_{l'mk\sigma}\ll~\hat{n}_{l\sigma'}f_{l'\sigma};f^{\dag}_{m\sigma}\gg\label{eq:10b}
\end{eqnarray}
\begin{eqnarray}
(\omega+\mu+\varepsilon_{fm\sigma'}-\varepsilon_{mk\sigma'}-\varepsilon_{fm\sigma})G^{m}_{fcf}
&=&\langle f^{\dag}_{m\sigma'}c_{mk\sigma'}\rangle+V^{\ast}_{mmk\sigma'}\ll~\hat{n}_{m\sigma'}f_{m\sigma};f^{\dag}_{m\sigma}\gg
\nonumber\\
&+&\sum_{k'}\big(V^{\ast}_{mk'\sigma}\ll~f^{\dag}_{m\sigma'}c_{mk\sigma'}c_{mk'\sigma};f^{\dag}_{m\sigma}\gg
-V_{mk'\sigma'}\ll~c^{\dag}_{mk'\sigma'}c_{mk\sigma'}f_{m\sigma};f^{\dag}_{m\sigma}\gg\big)\nonumber\\
&+&\sum_{l,l\neq m} V^{\ast}_{lmk\sigma'}\ll~f^{\dag}_{m\sigma'}f_{l\sigma'}f_{m\sigma};f^{\dag}_{m\sigma}\gg\label{eq:11}
\end{eqnarray}
\begin{eqnarray}
(\omega+\mu+\varepsilon_{fl\sigma}-\varepsilon_{lk\sigma}-\varepsilon_{fm\sigma})G^{lm}_{f\sigma cf}
&=&\langle f^{\dag}_{l\sigma}c_{lk\sigma}\rangle+V^{\ast}_{llk\sigma}\ll~\hat{n}_{l\sigma}f_{m\sigma};f^{\dag}_{m\sigma}\gg
+\sum_{k'}\big(V^{\ast}_{mk'\sigma}\ll~f^{\dag}_{m\sigma}c_{mk\sigma}c_{mk'\sigma};f^{\dag}_{m\sigma}\gg\nonumber\\
&-&V_{lk'\sigma}\ll~c^{\dag}_{lk'\sigma}c_{lk\sigma}f_{m\sigma};f^{\dag}_{m\sigma}\gg\big)
+\sum_{l',l'\neq l} V^{\ast}_{l'lk\sigma}\ll~f^{\dag}_{l\sigma}f_{l'\sigma}f_{m\sigma};f^{\dag}_{m\sigma}\gg\label{eq:11a}
\end{eqnarray}
\begin{eqnarray}
(\omega+\mu+\varepsilon_{fl\sigma'}-\varepsilon_{lk\sigma'}-\varepsilon_{fm\sigma})G^{lm}_{f\sigma' cf}
&=&\langle f^{\dag}_{l\sigma'}c_{lk\sigma'}\rangle+V^{\ast}_{llk\sigma'}\ll~\hat{n}_{l\sigma'}f_{m\sigma};f^{\dag}_{m\sigma}\gg
+\sum_{k'}\big(V^{\ast}_{mk'\sigma}\ll~f^{\dag}_{m\sigma'}c_{mk\sigma'}c_{mk'\sigma};f^{\dag}_{m\sigma}\gg\nonumber\\
&-&V_{lk'\sigma'}\ll~c^{\dag}_{lk'\sigma'}c_{lk\sigma'}f_{m\sigma};f^{\dag}_{m\sigma}\gg\big)
+\sum_{l',l'\neq l} V^{\ast}_{l'lk\sigma'}\ll~f^{\dag}_{l\sigma'}f_{l'\sigma'}f_{m\sigma};f^{\dag}_{m\sigma}\gg\label{eq:11b}
\end{eqnarray}
\begin{eqnarray}
(\omega+\mu+\varepsilon_{mk\sigma'}-\varepsilon_{fm\sigma'}-\varepsilon_{fm\sigma})G^{m}_{cff}
&=&\langle c^{\dag}_{mk\sigma'}f_{m\sigma'}\rangle+U_{mm}\ll~c^{\dag}_{mk\sigma'}f_{m\sigma'}f_{m\sigma};f^{\dag}_{m\sigma}\gg\nonumber\\
&+&2\sum_l\big(U_{lm\sigma\sigma}\ll~\hat{n}_{l\sigma}c^{\dag}_{mk\sigma'}f_{m\sigma'}f_{m\sigma};f^{\dag}_{m\sigma}\gg+U_{lm\sigma'\sigma}\ll
\hat{n}_{l\sigma'}c^{\dag}_{mk\sigma'}f_{m\sigma'}f_{m\sigma};f^{\dag}_{m\sigma}\gg\big)\nonumber\\
&-&V_{mmk\sigma'}\ll~\hat{n}_{m\sigma'}f_{m\sigma};f^{\dag}_{m\sigma}\gg
+\sum_{k'}\big(V^{\ast}_{mk'\sigma'}\ll~c^{\dag}_{mk\sigma'}c_{mk'\sigma'}f_{m\sigma};f^{\dag}_{m\sigma}\gg\nonumber\\
&+&V^{\ast}_{mk'\sigma'}\ll~c^{\dag}_{mk\sigma'}f_{m\sigma'}c_{mk'\sigma};f^{\dag}_{m\sigma}\gg\big)
-\sum_{l,l\neq m} V_{lmk\sigma'}\ll~f^{\dag}_{l\sigma'}f_{m\sigma'}f_{m\sigma};f^{\dag}_{m\sigma}\gg,\label{eq:12}
\end{eqnarray}
\begin{eqnarray}
(\omega+\mu+\varepsilon_{lk\sigma}-\varepsilon_{fl\sigma}-\varepsilon_{fm\sigma})G^{lm}_{cf\sigma f}
&=&\langle c^{\dag}_{lk\sigma}f_{l\sigma}\rangle+U_{lm\sigma\sigma}\ll~c^{\dag}_{lk\sigma}f_{l\sigma}f_{m\sigma};f^{\dag}_{m\sigma}\gg\nonumber\\
&+&\big(U_{mm}\ll~\hat{n}_{m\sigma'}c^{\dag}_{lk\sigma}f_{l\sigma}f_{m\sigma};f^{\dag}_{m\sigma}\gg+U_{ml\sigma'\sigma}\ll
\hat{n}_{m\sigma'}c^{\dag}_{lk\sigma}f_{l\sigma}f_{m\sigma};f^{\dag}_{m\sigma}\gg\big)\nonumber\\
&+&\big(U_{ll}\ll~\hat{n}_{l\sigma'}c^{\dag}_{lk\sigma}f_{l\sigma}f_{m\sigma};f^{\dag}_{m\sigma}\gg+U_{lm\sigma'\sigma}\ll
\hat{n}_{l\sigma'}c^{\dag}_{lk\sigma}f_{l\sigma}f_{m\sigma};f^{\dag}_{m\sigma}\gg\big)\nonumber\\
&+&\sum_{\substack{l'\neq l\\ l'\neq m}}\big(U_{l'm\sigma\sigma}\ll~\hat{n}_{l'\sigma}c^{\dag}_{lk\sigma}f_{l\sigma}f_{m\sigma};f^{\dag}_{m\sigma}\gg+U_{l'm\sigma'\sigma}\ll
\hat{n}_{l\sigma'}c^{\dag}_{lk\sigma}f_{l\sigma}f_{m\sigma};f^{\dag}_{m\sigma}\gg\nonumber\\
&&~~~~~+U_{l'l\sigma\sigma}\ll~\hat{n}_{l'\sigma}c^{\dag}_{lk\sigma}f_{l\sigma}f_{m\sigma};f^{\dag}_{m\sigma}\gg+U_{l'l\sigma'\sigma}\ll
\hat{n}_{l\sigma'}c^{\dag}_{lk\sigma}f_{l\sigma}f_{m\sigma};f^{\dag}_{m\sigma}\gg\big)\nonumber\\
&-&V_{llk\sigma}\ll~\hat{n}_{l\sigma}f_{m\sigma};f^{\dag}_{m\sigma}\gg
+\sum_{k'}\big(V^{\ast}_{lk'\sigma}\ll~c^{\dag}_{lk\sigma}c_{lk'\sigma}f_{m\sigma};f^{\dag}_{m\sigma}\gg\nonumber\\
&+&V^{\ast}_{mk'\sigma}\ll~c^{\dag}_{lk\sigma}f_{l\sigma}c_{mk'\sigma};f^{\dag}_{m\sigma}\gg\big)
-\sum_{l',l'\neq l} V_{l'lk\sigma}\ll~f^{\dag}_{l'\sigma}f_{l\sigma}f_{m\sigma};f^{\dag}_{m\sigma}\gg,\label{eq:12a}
\end{eqnarray}
\begin{eqnarray}
(\omega+\mu+\varepsilon_{lk\sigma'}-\varepsilon_{fl\sigma'}-\varepsilon_{fm\sigma})G^{lm}_{cf\sigma' f}
&=&\langle c^{\dag}_{lk\sigma'}f_{l\sigma'}\rangle+U_{lm\sigma'\sigma}\ll~c^{\dag}_{lk\sigma'}f_{l\sigma'}f_{m\sigma};f^{\dag}_{m\sigma}\gg\nonumber\\
&+&\big(U_{mm}\ll~\hat{n}_{m\sigma'}c^{\dag}_{lk\sigma'}f_{l\sigma'}f_{m\sigma};f^{\dag}_{m\sigma}\gg+U_{ml\sigma'\sigma'}\ll
\hat{n}_{m\sigma'}c^{\dag}_{lk\sigma'}f_{l\sigma'}f_{m\sigma};f^{\dag}_{m\sigma}\gg\big)\nonumber\\
&+&\big(U_{ll}\ll~\hat{n}_{l\sigma}c^{\dag}_{lk\sigma'}f_{l\sigma'}f_{m\sigma};f^{\dag}_{m\sigma}\gg+U_{lm\sigma\sigma}\ll
\hat{n}_{l\sigma}c^{\dag}_{lk\sigma'}f_{l\sigma'}f_{m\sigma};f^{\dag}_{m\sigma}\gg\big)\nonumber\\
&+&\sum_{\substack{l'\neq l\\ l'\neq m}}\big(U_{l'm\sigma\sigma}\ll~\hat{n}_{l'\sigma}c^{\dag}_{lk\sigma'}f_{l\sigma'}f_{m\sigma};f^{\dag}_{m\sigma}\gg+U_{l'm\sigma'\sigma}\ll
\hat{n}_{l\sigma'}c^{\dag}_{lk\sigma'}f_{l\sigma'}f_{m\sigma};f^{\dag}_{m\sigma}\gg\nonumber\\
&&~~~~~+U_{l'l\sigma\sigma'}\ll~\hat{n}_{l'\sigma}c^{\dag}_{lk\sigma'}f_{l\sigma'}f_{m\sigma};f^{\dag}_{m\sigma}\gg+U_{l'l\sigma'\sigma'}\ll
\hat{n}_{l\sigma'}c^{\dag}_{lk\sigma'}f_{l\sigma'}f_{m\sigma};f^{\dag}_{m\sigma}\gg\big)\nonumber\\
&-&V_{llk\sigma'}\ll~\hat{n}_{l\sigma'}f_{m\sigma};f^{\dag}_{m\sigma}\gg
+\sum_{k'}\big(V^{\ast}_{lk'\sigma'}\ll~c^{\dag}_{lk\sigma'}c_{lk'\sigma'}f_{m\sigma};f^{\dag}_{m\sigma}\gg\nonumber\\
&+&V^{\ast}_{mk'\sigma}\ll~c^{\dag}_{lk\sigma'}f_{l\sigma'}c_{mk'\sigma};f^{\dag}_{m\sigma}\gg\big)
-\sum_{l',l'\neq l} V_{l'lk\sigma'}\ll~f^{\dag}_{l'\sigma'}f_{l\sigma'}f_{m\sigma};f^{\dag}_{m\sigma}\gg,\label{eq:12b}
\end{eqnarray}
where the Coulomb interaction strength will change accordingly for the multi-orbital case due to the pre-existence of charge, and we have used the following abbreviations for the Green's functions on the left hand side:
\begin{eqnarray}
G^{m}_{nf}=\ll~\hat{n}_{m\sigma'}f_{m\sigma};f^{\dag}_{m\sigma}\gg,&\qquad&G^{m}_{fcf}=\ll~f^{\dag}_{m\sigma'}c_{mk\sigma'}f_{m\sigma};f^{\dag}_{m\sigma}\gg,\nonumber\\
G^{m}_{nc}=\ll~\hat{n}_{m\sigma'}c_{mk\sigma};f^{\dag}_{m\sigma}\gg,&\qquad&G^{m}_{cff}=\ll~c^{\dag}_{mk\sigma'}f_{m\sigma'}f_{m\sigma};f^{\dag}_{m\sigma}\gg.\nonumber\\
G^{lm}_{n\sigma f}=\ll~\hat{n}_{l\sigma}f_{m\sigma};f^{\dag}_{m\sigma}\gg,&\qquad&G^{lm}_{f\sigma cf}=\ll~f^{\dag}_{l\sigma}c_{lk\sigma}f_{m\sigma};f^{\dag}_{m\sigma}\gg,\nonumber\\
G^{lm}_{n\sigma c}=\ll~\hat{n}_{l\sigma}c_{mk\sigma};f^{\dag}_{m\sigma}\gg,&\qquad&G^{lm}_{cf\sigma f}=\ll~c^{\dag}_{lk\sigma}f_{l\sigma}f_{m\sigma};f^{\dag}_{m\sigma}\gg.\nonumber\\
G^{lm}_{n\sigma' f}=\ll~\hat{n}_{l\sigma'}f_{m\sigma};f^{\dag}_{m\sigma}\gg,&\qquad&G^{lm}_{f\sigma' cf}=\ll~f^{\dag}_{l\sigma'}c_{lk\sigma'}f_{m\sigma};f^{\dag}_{m\sigma}\gg,\nonumber\\
G^{lm}_{n\sigma' c}=\ll~\hat{n}_{l\sigma'}c_{mk\sigma};f^{\dag}_{m\sigma}\gg,&\qquad&G^{lm}_{cf\sigma' f}=\ll~c^{\dag}_{lk\sigma'}f_{l\sigma'}f_{m\sigma};f^{\dag}_{m\sigma}\gg.\nonumber
\end{eqnarray}
In the above equations of motion, we will drop the Green's functions involving odd number of operators of one arbitrary orbital, e.g. $\ll~f^{\dag}_{l'\sigma'}f_{l\sigma'}f_{m\sigma};f^{\dag}_{m\sigma}\gg$ etc., and attribute them to the effects of inter-orbital hybridizations which will be taken into account through the DMFT self-consistency conditions.

For the three-particle Green's functions we used the following decouplings,
\begin{eqnarray}
\ll~\hat{n}_{l\sigma}\hat{n}_{m\sigma'}f_{m\sigma};f^{\dag}_{m\sigma}\gg\approx\bar{n}_{l\sigma}\ll~\hat{n}_{m\sigma'}f_{m\sigma};f^{\dag}_{m\sigma}\gg,\\
\ll~\hat{n}_{l\sigma'}\hat{n}_{m\sigma'}f_{m\sigma};f^{\dag}_{m\sigma}\gg\approx\bar{n}_{l\sigma'}\ll~\hat{n}_{m\sigma'}f_{m\sigma};f^{\dag}_{m\sigma}\gg,\\
\ll~\hat{n}_{m\sigma'}\hat{n}_{l\sigma}f_{m\sigma};f^{\dag}_{m\sigma}\gg\approx\bar{n}_{m\sigma'}\ll~\hat{n}_{l\sigma}f_{m\sigma};f^{\dag}_{m\sigma}\gg,\\
\ll~\hat{n}_{l\sigma'}\hat{n}_{l\sigma}f_{m\sigma};f^{\dag}_{m\sigma}\gg\approx\bar{n}_{l\sigma'}\ll~\hat{n}_{l\sigma}f_{m\sigma};f^{\dag}_{m\sigma}\gg,\\
\ll~\hat{n}_{m\sigma'}\hat{n}_{l\sigma'}f_{m\sigma};f^{\dag}_{m\sigma}\gg\approx\bar{n}_{m\sigma'}\ll~\hat{n}_{l\sigma'}f_{m\sigma};f^{\dag}_{m\sigma}\gg,\\
\ll~\hat{n}_{l\sigma}\hat{n}_{l\sigma'}f_{m\sigma};f^{\dag}_{m\sigma}\gg\approx\bar{n}_{l\sigma}\ll~\hat{n}_{l\sigma'}f_{m\sigma};f^{\dag}_{m\sigma}\gg.
\end{eqnarray}
\end{widetext}


\begin{thebibliography}{99}
\bibitem {code} This MO-EOM impurity solver code will be made available to the general public in the future.
\end{thebibliography}
\end{document}